\documentclass[preprint,12pt]{elsarticle}

\usepackage{amssymb}
\journal{}

\begin{document}

\begin{frontmatter}



\title{Left dihedral codes over Galois rings ${\rm GR}(p^2,m)$}


\author{Yonglin Cao$^{a \ \ast}$,  Yuan Cao$^{b}$, Fang-Wei Fu$^{c}$}

\address{$^{a}$School of Sciences,
Shandong University of Technology, Zibo, Shandong 255091, China
\vskip 1mm $^{b}$College of Mathematics and Econometrics, Hunan University, Changsha 410082, China
 \vskip 1mm $^{c}$Chern Institute of Mathematics and LPMC, Nankai University, Tianjin 300071, China}
 \cortext[cor1]{corresponding author.   \\
E-mail addresses: ylcao@sdut.edu.cn (Yonglin Cao), \ yuan$_{-}$cao@hnu.edu.cn (Yuan Cao), \ fwfu@nankai.edu.cn (F.--W. Fu).}

\begin{abstract}
 Let $D_{2n}=\langle x,y\mid x^n=1, y^2=1, yxy=x^{-1}\rangle$ be a dihedral group, and $R={\rm GR}(p^2,m)$ be a Galois ring
of characteristic $p^2$ and cardinality $p^{2m}$ where $p$ is a prime. Left ideals of the group ring $R[D_{2n}]$ are called
left dihedral codes over $R$ of length $2n$, and abbreviated as left $D_{2n}$-codes over $R$.
Let ${\rm gcd}(n,p)=1$ in this paper. Then any left $D_{2n}$-code
over $R$ is uniquely decomposed into a direct sum of concatenated codes with inner codes ${\cal A}_i$ and outer codes $C_i$, where ${\cal A}_i$ is a cyclic code over $R$ of length $n$ and $C_i$ is a skew cyclic code of length $2$ over an extension Galois ring or principal ideal ring of $R$, and a generator matrix and basic parameters for each outer code $C_i$ is given. Moreover, a formula to count the number of these codes is obtained, the dual
code for each left $D_{2n}$-code is determined and all self-dual left $D_{2n}$-codes and self-orthogonal left $D_{2n}$-codes over $R$ are presented, respectively.
\end{abstract}


\end{frontmatter}


\section{Introduction}
\label{intro} \noindent
After the celebrated results in the 1990¡¯s ([14], [24], [31]) that many important yet
seemingly non-linear codes over finite fields are actually closely related to linear codes over the ring of integers modulo
four, codes over $\mathbb{Z}_4$ in particular, and codes over finite commutative chain rings in general, have received a great
deal of attention. Examples of finite commutative chain rings include the ring $\mathbb{Z}_{p^k}$ of integers modulo $p^k$ for a prime $p$, and the Galois rings ${\rm GR}(p^k, m)$, i.e. the Galois extension of degree $m$ of $\mathbb{Z}_{p^k}$. These classes of rings have been used widely as an alphabet for codes.

\par
  Let $R$ be a finite chain ring with identity, $R^\times$ be the multiplicative group of invertible elements of $R$, $N$ a positive integer and
$R^N=\{(r_1,\ldots,r_N)$ $\mid r_j\in R, \ j=1,\ldots,N\}$ be the free $R$-module of rank $N$ with the coordinate component addition
and scalar multiplication by elements of $R$. Then \textit{linear codes} over $R$ (or  \textit{linear $R$-codes}) of length $N$ are defined as $R$-submodules
of $R^N$. For any fixed $\lambda\in R^\times$, a linear $R$-code $\mathcal{C}$ of length $N$ is said to be
\textit{$\lambda$-constacyclic} if $(\lambda c_{n-1},c_0,c_1,\ldots,c_{n-2})\in \mathcal{C}$ for all $(c_0,c_1,\ldots,c_{n-1})\in \mathcal{C}$, and $\mathcal{C}$ is said to be
\textit{cyclic} (\textit{negacyclic}) when $\lambda=1$ ($\lambda=-1$).
Readers are referred to [17], [20], [21], [25], [27], [32] and [34] for results on linear codes, cyclic codes and constacyclic codes over finite
chain rings. Moreover, various decoding schemes for codes over Galois rings have been considered in [11--13].

\par
  Now, let $N=ln$. A linear code $\mathcal{C}$ over $R$ of length $N$ is called a \textit{$l$-quasi-cyclic code} of length $ln$ if
$$(c_{1,n-1},c_{1,0},c_{1,1},\ldots,c_{1,n-2},\ldots,c_{l,n-1},c_{l,0},c_{l,1}\ldots,c_{l,n-2})\in \mathcal{C}$$
 for
all $(c_{1,0},c_{1,1},\ldots,c_{1,n-1},\ldots,c_{l,0},c_{l,1},\ldots,c_{l,n-1})\in \mathcal{C}$. Quasi-cyclic codes and their generalizations over
finite chain rings have been received a great
deal of attention. For example, see [1], [4], [15], [16] and [19].

\par
  In this paper, let
$$D_{2n}=\langle x,y\mid x^n=1, y^2=1, yxy=x^{-1}\rangle=\{x^iy^j\mid 0\leq i\leq n-1, j=0,1\}$$
 be a dihedral group of order $n$.
The group ring $R[D_{2n}]$ is a free $R$-module with basis $D_{2n}$. Addition, multiplication with
scalars $c\in R$ and multiplication are defined by: for any $a_g,b_g\in R$ where $g\in D_{2n}$,
$$\sum_{g\in D_{2n}}a_g g+\sum_{g\in D_{2n}}b_g g=\sum_{g\in D_{2n}}(a_g+b_g) g, \
c(\sum_{g\in D_{2n}}a_g g)=\sum_{g\in D_{2n}}ca_g g,$$
$$(\sum_{g\in D_{2n}}a_g g)(\sum_{g\in D_{2n}}b_g g)=\sum_{g\in D_{2n}}(\sum_{uv=g}a_ub_v)g.$$
Then $R[D_{2n}]$ is a noncommutative ring with identity $1=1_{R}1_{D_{2n}}$ where
$1_{R}$ and $1_{D_{2n}}$ is the identity elements of $R$ and $D_{2n}$ respectively. Readers are referred to [33]
for more details on group rings.

\par
   For any $a=(a_{0,0}, a_{1,0},\ldots, a_{n-1,0},a_{0,1}, a_{1,1},\ldots, a_{n-1,1})\in R^{2n}$, we define
$$\Psi(a)=\sum_{i=0}^{n-1}a_{i,0}x^i+\sum_{i=0}^{n-1}a_{i,1}x^iy.$$
Then $\Psi$ is an $R$-module isomorphism from $R^{2n}$ onto
$R[D_{2n}]$. As a natural generalization of [23], a nonempty subset $\mathcal{C}$ of the $R$-module $R^{2n}$
is called a \textit{left dihedral code} (or \textit{left $D_{2n}$-code} for more clear) over $R$ if $\Psi(\mathcal{C})$ is a left ideal of  $R[D_{2n}]$. As usual, we will identify
$\mathcal{C}$ with $\Psi(\mathcal{C})$ in this paper. Since
$$\Psi(a_{n-1,0}, a_{0,0},a_{1,0},\ldots, a_{n-2,0}, a_{n-1,1},a_{0,1},a_{1,1},\ldots, a_{n-2,1})=x\Psi(a)$$
for all $a=(a_{0,0}, a_{1,0},\ldots, a_{n-1,0},a_{0,1}, a_{1,1},\ldots, a_{n-1,1})\in R^{2n}$, we see that
every left $D_{2n}$-code over $R$ is a $2$-quasi-cyclic code over $R$ of length $2n$.

\par
  When $R=\mathbb{F}_q$ is a finite field of cardinality $q$, there have been many research results
on codes as two-sided ideals and left ideals in a finite group algebra over finite fields. For example, Dutra et al [22] investigated codes that are given as two-sided ideals in a semisimple finite group algebra
$\mathbb{F}_q[G]$ defined by idempotents constructed from subgroups of a finite group $G$,
and given a criterion to decide when these ideals are all the minimal two-sided ideals of $\mathbb{F}_q[G]$ in the case
when $G$ is a dihedral group. McLoughlin [30]
provided a new construction of the self-dual, doubly-even and extremal [48,24,12]
binary linear block code using a
zero divisor in the group ring $\mathbb{F}_2[D_{48}]$ where $D_{48}$ is a dihedral group of order $24$.

\par
   Recently,  Brochero Mart\'{i}nez [10] shown all central irreducible
idempotents and their Wedderburn decomposition of the
dihedral group algebra $\mathbb{F}_q[D_{2n}]$, in the case when every divisor
of $n$ divides $q-1$. This characterization depends to the relation of the irreducible
idempotents of the cyclic group algebra $\mathbb{F}_q[C_{n}]$ and the central
irreducible idempotents of the group algebras $\mathbb{F}_q[D_{2n}]$.
Gabriela and Inneke [23] provided algorithms to construct minimal left
group codes. These are based on results describing a complete set of orthogonal primitive idempotents in each
Wedderburn component of a semisimple finite group algebra $\mathbb{F}_q[G]$ for a large class of groups $G$.

\par
  More importantly, Bazzi and Mitter [3] shown that for infinitely
many block lengths a random left ideal in the binary group algebra of
the dihedral group is an asymptotically good rate-half code with
a high probability. Mart\'{i}nez-P\'{e}rez and Willems [29] proved
that the class of binary
self-dual doubly even $2$-quasi-cyclic transitive codes is asymptotically
good. A system theory for left $D_{2n}$-codes over finite fields was given [18].

\par
   Let $R={\rm GR}(p^2,m)$ be a Galois ring  of characteristic $p^2$ and cardinality $p^{2m}$ and
${\rm gcd}(n,p)=1$, we try to achieve the following two goals:

\par
   $\diamondsuit$ Develop a system theory for left $D_{2n}$-codes over $R$ using an elementary method. Precisely, only finite field theory,
Galois ring theory and basic theory of cyclic codes and skew
cyclic codes are used, and it does not involve any group algebra language and technique except the definition for left $D_{2n}$-codes.

\par
   $\diamondsuit$ Provide a precise expression for all distinct left $D_{2n}$-codes, their dual codes, self-orthogonal and self-dual left $D_{2n}$-codes over $R$.

\noindent
   Using the
expression provided, one can present left $D_{2n}$-codes, self-orthogonal and self-dual left $D_{2n}$-codes over $R$ for
specific $n$, $m$ and $p$ (not too big) conveniently and easily, and design left $D_{2n}$-codes for their requirements
directly.

\vskip3mm \par
   The present paper is organized as follows. In section 2,
we present every left $D_{2n}$-code and its dual code over $R$ by their direct sum decomposition, where
each direct summand is a concatenated code, respectively. In particular, we determine the inner code and give a generator matrix and basic parameters for the outer code of each concatenated code precisely. As a corollary, we obtain a formula to count the number of all left $D_{2n}$-codes over $R$.
 In section 3, we give a proof for the main results (Theorem 2.5) of this paper by use of the known results for left dihedral codes over
 finite fields and Galois ring theory. Then
 we present all self-dual and all self-orthogonal left $D_{2n}$-codes over $R$ in Section 4. Finally, we list 60 optimal $\mathbb{Z}_4$-linear codes of length $30$ obtained from
left $D_{30}$-codes over $\mathbb{Z}_4$.

%%%%%%%%%%%%%%%%%%%%%%%%%%%%%%%%%%%%%%%%%%%%%%%%%%%%%%%%%%%%%%%%%%%%%%%%%%%%%%%%%%%%

%%%%%%%%%%%%%%%%%%%%%%%%%%%%%%%%%%%%%%%%%%%%%%%%%%%%%%%%%%%%%%%%%%%%%%%%%%%%%%%%%%%%
\section{Concatenated structure of left $D_{2n}$-codes over $R$}
\noindent
  In this section, we give the concatenated structure of every left $D_{2n}$-code and its Euclidian dual code over $R$ explicitly, and obtain a formula to count the number of all left $D_{2n}$-codes over $R$ precisely.
In this paper, we adopt the following notations.

\par
   Let $R={\rm GR}(p^2,m)$ be a fixed Galois ring
of characteristic $p^2$ and cardinality $p^{2m}$, $\mathbb{F}_{p^m}=R/pR$ be the residue class field
of $R$ modulo its unique maximal ideal $pR$ and $^{-}: R\rightarrow \mathbb{F}_{p^m}$ be the natural
surjective homomorphism of rings from $R$ onto $\mathbb{F}_{p^m}$ defined by $\overline{a}=a$ (mod $p$) for
all $a\in R$.

\vskip 3mm\noindent
  {\bf Example} Let $\mathbb{Z}_{p^2}=\{0,1,\ldots,p^2-1\}$ be the ring of integers modulo $p^2$,
$h(x)=x^m+h_{m-1}x^{m-1}+\ldots+h_1x+h_0$ be a fixed
basic irreducible monic polynomial of degree $m$ in $\mathbb{Z}_{p^2}[x]$, i.e. $h(x)$ is a monic polynomial of degree $m$ in $\mathbb{Z}_{p^2}[x]$
such that $h(x)$ (mod $p$) is irreducible in $\mathbb{Z}_p[x]$, and let
$$R=\{a_0+a_1z+\ldots+a_{m-1}z^{m-1}\mid a_0,a_1,\ldots,a_{m-1}\in \mathbb{Z}_{p^2}\}=\mathbb{Z}_{p^2}[z]$$
in which $z^m=-h_0-h_1z-\ldots-h_{m-1}z^{m-1}$.
Then $R$ is a Galois ring
of characteristic $p^2$ and cardinality $p^{2m}$. Denote $\overline{a}=a$ (mod $p$) for any $a\in \mathbb{Z}_{p^2}$.
It is known that $\mathbb{F}_{p^m}=\mathbb{Z}_{p}[x]/\langle \overline{h}(x)\rangle=\mathbb{Z}_{p}[\overline{z}]$, where $\overline{h}(x)=x^m+\overline{h}_{m-1}x^{m-1}+\ldots+\overline{h}_1x+\overline{h}_0\in \mathbb{Z}_{p}[x]$ and
$\overline{z}^m=-\overline{h}_0-\overline{h}_1\overline{z}-\ldots-\overline{h}_{m-1}\overline{z}^{m-1}$, and the natural surjective homomorphism
of rings $^{-}: R\rightarrow \mathbb{F}_{p^m}$ is given by
$$a_0+a_1z+\ldots+a_{m-1}z^{m-1}\mapsto \overline{a}_0+\overline{a}_1\overline{z}+\ldots+\overline{a}_{m-1}\overline{z}^{m-1}$$
for any $a_0,a_1,\ldots,a_{m-1}\in \mathbb{Z}_{p^2}$ (cf. [36] Theorem 14.1).

\vskip 3mm\par
The ring homomorphism $^{-}: R\rightarrow \mathbb{F}_{p^m}$ can be extended to a surjective homomorphism of polynomial rings
from $R[x]$ onto $\mathbb{F}_{p^m}[x]$ in the natural way:
$$g(x)\mapsto \overline{g}(x)=\sum_{i=0}^k\overline{g}_ix^i, \ \forall g(x)=\sum_{i=0}^kg_ix^i\in R[x] \
{\rm with} \ g_i\in R.$$
We will denote this homomorphism by $^{-}$ as well.
Then a monic polynomial $g(x)\in R[x]$ is said to be \textit{basic irreducible} (\textit{basic primitive}) if $\overline{g}(x)$ is an irreducible (primitive) polynomial in $\mathbb{F}_{p^m}[x]$. Denote
$$\mathcal{A}=R[x]/\langle x^n-1\rangle=\{a(x)\mid a(x)=\sum_{j=0}^{n-1}a_jx^j, \ a_0,a_1,\ldots,a_{n-1}\in R\}$$
where the arithmetic is done modulo $x^n-1$. For any positive integer $l$, by a \textit{linear $\mathcal{A}$-code} of length $l$ we mean an
$\mathcal{A}$-submodule of $\mathcal{A}^l=\{(\alpha_1,\ldots,\alpha_l)$ $\mid \alpha_1,\ldots,\alpha_l\in \mathcal{A}\}$.

\par
  If $l=1$, linear $\mathcal{A}$-codes are exactly ideals of the ring $\mathcal{A}$ which are the well-known \textit{cyclic codes}
over $R$ of length $n$ (cf. [37] Proposition 1.1). If $l\geq 2$, every linear $\mathcal{A}$-code of $l$ can be seen as an $R$-linear code of length $nl$ by
replacing each $a(x)=\sum_{j=0}^{n-1}a_jx^j\in \mathcal{A}$ with $(a_0,a_1,\ldots,a_{n-1})\in R^n$, and these $R$-linear codes of length $nl$
are the well known \textit{$l$-quasi-cyclic codes}
over $R$ of length $nl$.

\par
   Let $\Omega_n=\{1,x,\ldots,x^{n-1}\}$ be the cyclic subgroup of $D_{2n}$ generated by
$x$. Then we can identify the group ring $R[\Omega_n]$ with $\mathcal{A}$
in the natural way. Hence every element of the group ring $R[D_{2n}]$ is uniquely expressed as:
$a_0(x)+a_1(x)y$, $a_0(x), a_1(x)\in \mathcal{A}.$
Now, we define a map $\Xi: \mathcal{A}^2\rightarrow R[D_{2n}]$ by
$$(a_0(x), a_1(x))\mapsto a_0(x)+a_1(x)y \ (\forall a_0(x), a_1(x)\in \mathcal{A}).$$
It is routine to check that $\Xi$ is an isomorphism of $\mathcal{A}$-modules from $\mathcal{A}^2$ onto $R[D_{2n}]$.
As stated above, $\mathcal{A}$-submodules of $\mathcal{A}^2$ are $2$-quasi-cyclic codes over $R$ of length $2n$. We will identify
$R[D_{2n}]$ with $\mathcal{A}^2$ under $\Xi$ in this paper. Then by
$yx=x^{-1}y$ in $D_{2n}$, we deduce the following conclusion.

\vskip 3mm \noindent
  {\bf Theorem 2.1} \textit{Let $\emptyset\neq \mathcal{C}\subseteq \mathcal{A}^2$. Then $\mathcal{C}$ is a left ideal
of $R[D_{2n}]$ if and only if $\mathcal{C}$ is an $\mathcal{A}$-submodule of $\mathcal{A}^2$ satisfying the following condition}
$$(a_1(x^{-1}),a_0(x^{-1}))\in \mathcal{C}, \ \forall (a_0(x),a_1(x))\in \mathcal{C}.$$

\par
   For any nonzero polynomial $g(x)=\sum_{i=0}^da_ix^i\in R[x]$ of degree $d$, recall that the \textit{reciprocal polynomial}
of $g(x)$ is defined by
$$g^\ast(x)=(g(x))^\ast=x^dg(\frac{1}{x})=a_d+a_{d-1}x+\ldots+a_0x^d.$$
$g(x)$ is said to be \textit{self-reciprocal}
if $g^\ast(x)=ug(x)$ for some $u\in R^{\times}$.

\par
  As ${\rm gcd}(p,n)=1$, we have
$x^n-1=\prod_{i=0}^{r+t}f_i(x)$,
where $f_0(x),f_1(x),\ldots$, $f_{r+t}(x)$ are pairwise coprime monic polynomials in $R[x]$ such that

\vskip 2mm\par
  $\bullet$ $f_0(x)=x-1$,

\vskip 2mm\par
  $\bullet$  $f_i(x)$ is a self-reciprocal and basic irreducible polynomial of degree $d_i$, $i=1,\ldots,r$,

\vskip 2mm\par
  $\bullet$ $f_{i}(x)=\rho_i(x)\rho_i^\ast(x)$ where $\rho_i(x)$ is a monic basic irreducible polynomial of degree $d_i$ such that
$\rho_i(x)$ and $\rho_i^\ast(x)$ are coprime in $R[x]$, $i=r+1,\ldots,r+t$.

\vskip 2mm\noindent
Then it is clear that $n=\sum_{i=0}^rd_i+2\sum_{i=r+1}^{r+t}d_i$ where $d_0=1$.
Denote

\vskip 2mm\par
 $\bullet$ $A_i=R[x]/\langle f_i(x)\rangle$ where we consider elements of $A_i$ as polynomials
in $R[x]$ of degree $<{\rm deg}(f_i(x))$ and the arithmetic is done modulo $f_i(x)$, for all $i=0,1,\ldots,r+t$.

\vskip 2mm\par
 $\bullet$ $\Upsilon_{i,1}=R[x]/\langle \rho_i(x)\rangle$ and $\Upsilon_{i,2}=R[x]/\langle \rho_i^\ast(x)\rangle$
where we consider elements of $\Upsilon_{i,1}$ (resp. $\Upsilon_{i,2}$) as polynomials
in $R[x]$ of degree $<d_i$ and the arithmetic is done modulo $\rho_i(x)$ (resp. $\rho_i^\ast(x)$), for $i=r+1,\ldots,r+t$.

\vskip 2mm\noindent
As a direct corollary of Wan [36] Theorem 14.23 and Theorem 14.27, we have the following lemma.

\vskip 3mm \noindent
   {\bf Lemma 2.2} (i) \textit{For each integer $i$, $0\leq i\leq r$, $A_i$ is a Galois ring of characteristic $p^2$ and cardinality $p^{2md_i}$
which is a Galois extension of $R$ with degree $d_i$, and there is an invertible element $\zeta_i(x)$ in $A_i$ of multiplicative order
$p^{md_i}-1$}.

\par
   (ii) \textit{For each integer $i$, $r+1\leq i\leq r+t$, $\Upsilon_{i,1}$ is a Galois ring of characteristic $p^2$ and cardinality $p^{2md_i}$
which is a Galois extension of $R$ with degree $d_i$, and there is an invertible element $\zeta_i(x)$ in $\Upsilon_{i,1}$ of multiplicative order
$p^{md_i}-1$}.

\par
   (iii) \textit{$A_i$ is a free $R$-module of rank ${\rm deg}(f_i(x))$  with
an $R$-basis $\{1,x, \ldots$, $x^{{\rm deg}(f_i(x))-1}\}$ for any $0\leq i\leq r+t$}.

\vskip 3mm\par
   For each $0\leq i\leq r+t$, denote
$F_i(x)=\frac{x^n-1}{f_i(x)}\in R[x]$. Then $F_i(x)$ and $f_i(x)$ are coprime polynomials in $R[x]$. Hence there are
polynomials $u_i(x),v_i(x)\in R[x]$ such that
\begin{equation}
u_i(x)F_i(x)+v_i(x)f_i(x)=1.
\end{equation}
In the rest of this paper, let $\varepsilon_i(x)\in \mathcal{A}$ satisfying

\vskip 2mm \par
  $\bullet$ $\varepsilon_i(x)\equiv u_i(x)F_i(x)=1-v_i(x)f_i(x)$  (mod $x^n-1$).

\vskip 2mm \noindent
 Then from classical ring theory, we deduce the following
lemma.

\vskip 3mm \noindent
   {\bf Lemma 2.3} (cf. [37] Theorem 2.7 and its proof)
  (i) \textit{$\sum_{i=0}^{r+t}\varepsilon_i(x)=1$, $\varepsilon_i(x)^2=\varepsilon_i(x)$ and $\varepsilon_i(x)\varepsilon_j(x)=0$ for all $0\leq i\neq j\leq r+t$ in the ring ${\cal A}$}.

\vskip 2mm\par
  (ii) \textit{${\cal A}=\oplus_{i=0}^{r+t}{\cal A}_i$, where ${\cal A}_i=\varepsilon_i(x){\cal A}$
with $\varepsilon_i(x)$ as its multiplicative identity. Moreover, this decomposition is a ring direct
sum in that ${\cal A}_i{\cal A}_j=\{0\}$ for all $0\leq i\neq j\leq r+t$}.

\vskip 2mm\par
  (iii) \textit{For each $0\leq i\leq r+t$, let $A_i=R[x]/\langle f_i(x)\rangle$.
Then the map
$$\varphi_i: a(x)\mapsto \varepsilon_i(x)a(x) \ ({\rm mod} \ x^n-1), \ \forall a(x)\in A_i$$
is an isomorphism of rings from $A_i$ onto ${\cal A}_i$}.

\vskip 3mm \par
   As usual, we will identify each $(a_0,a_1,\ldots,a_{n-1})\in R^n$ with
$a_0+a_1x+\ldots+a_{n-1}x^{n-1}\in \mathcal{A}$. Then we have
the following properties for ${\cal A}_i=\varepsilon_i(x){\cal A}$.

\vskip 3mm \noindent
   {\bf Corollary 2.4} \textit{Let $0\leq i\leq r+t$. Then}

\par
   (i) (cf. [37] Proposition 4.3) \textit{${\cal A}_i$ is a cyclic code over $R$ of length $n$ with
parity check polynomial $f_i(x)$ and generating idempotent $\varepsilon_i(x)$}.

\par
   (ii) \textit{${\cal A}_i$ is a free $R$-submodule of $\mathcal{A}=R[x]/\langle x^n-1\rangle$ with
an $R$-basis $\{\varepsilon_i(x),x\varepsilon_i(x)$, $\ldots,x^{{\rm deg}(f_i(x))-1}\varepsilon_i(x)\}$. Hence
$|{\cal A}_i|=p^{2m\cdot{\rm deg}(f_i(x))}$}.

\vskip 3mm \noindent
   {\bf Proof.} (ii) As $f_i(x)$ is a monic divisor of $x^n-1$ in $R[x]$,
$A_i=R[x]/\langle f_i(x)\rangle$ is a free $R$-module with an $R$-basis $\{1,x,\ldots,x^{{\rm deg}(f_i(x))-1}\}$,
which implies $|A_i|=|R|^{{\rm deg}(f_i(x))}=p^{2m\cdot{\rm deg}(f_i(x))}$. From this and by
Lemma 2.3(iii), we see that $\{\varepsilon_i(x),x\varepsilon_i(x),\ldots,x^{{\rm deg}(f_i(x))-1}\varepsilon_i(x)\}$
is a free $R$-basis of ${\cal A}_i$.
\hfill $\Box$

\vskip 3mm\par
  Now, let $C_i$ be an $A_i$-linear code of length $2$, i.e., $C_i$ is an $A_i$-submodule of $A_i^2=\{(b_0(x),b_1(x))\mid b_0(x),b_1(x)\in A_i\}$.
For each $\xi=(b_0(x),b_1(x))\in A_i^2$, we denote by
${\rm w}_{H}^{(A_i)}(\xi)=|\{j\mid b_j(x)\neq 0 \ {\rm in} \ A_i, \ j=0,1\}|$ the Hamming weight of $\xi$. Then the
 minimum Hamming distance of $C_i$ is given by
$$d_{H}^{(A_i)}(C_i)={\rm min}\{{\rm w}_{H}^{(A_i)}(\xi) \mid \xi\neq 0, \ \xi\in C_i\}.$$
   As a natural generalization of the concept for concatenated codes over finite field (cf. [35, Definition 2.1]), we define the
\textit{concatenated code} of the inner code $\mathcal{A}_i$ and the outer code $C_i$ as following
\begin{eqnarray*}
\mathcal{A}_i\Box_{\varphi_i}C_i&=&\{(\varphi_i(b_0(x)),\varphi_i(b_1(x)))\mid (b_0(x),b_1(x))\in C_i\}\\
 &=&\{(\varepsilon_i(x)b_0(x),\varepsilon_i(x)b_1(x))\mid (b_0(x),b_1(x))\in C_i\}\subseteq \mathcal{A}_i^2.
\end{eqnarray*}
By Lemma 2.3(iii), we see that $\mathcal{A}_i\Box_{\varphi_i}C_i$ is a $2$-quasi-cyclic code over $R$ of length $2n$
with cardinality $|\mathcal{A}_i\Box_{\varphi_i}C_i|=|C_i|$. Moreover, the minimum Hamming distance of
$\mathcal{A}_i\Box_{\varphi_i}C_i$ satisfies
$$d_{H}^{(R)}(\mathcal{A}_i\Box_{\varphi_i}C_i)\geq d_{H}^{(R)}(\mathcal{A}_i)\cdot d_{H}^{(A_i)}(C_i)$$
where $d_{H}^{(R)}(\mathcal{A}_i)$ is the minimum Hamming weight of $\mathcal{A}_i$ as a linear $R$-code of
length $n$.

\par
  As the end of this section, we list all distinct
left $D_{2n}$-codes over $R$ and their Euclidean dual codes by the following theorem.

\vskip 3mm \noindent
  {\bf Theorem 2.5} \textit{Using the notations above, all distinct left $D_{2n}$-codes $\mathcal{C}$ over $R$ and
their Euclidean dual codes  $\mathcal{C}^{\bot_E}$ are given by}
$$\mathcal{C}=\oplus_{i=0}^{r+t}\mathcal{A}_i\Box_{\varphi_i}C_i
\ {\rm and} \ \mathcal{C}^{\bot_E}=\oplus_{i=0}^{r+t}\mathcal{A}_i\Box_{\varphi_i}V_i$$
\textit{respectively, where $C_i$ and $V_i$ are $A_i$-linear codes of length $2$ with
generator matrices $G_i$ and $H_i$ respectively, given by the one of following three tables}.

\par
  (I) \textit{Let $0\leq i\leq r$. We have one of the following two subcases}.

\par
  (i) \textit{Let $d_i=1$. Then}

\par
  (i-1) \textit{If $p$ is odd, $(G_i,H_i)$ is given by the following table}.

\begin{center}
\begin{tabular}{lcll|c}\hline
  $N_i$  &  $G_i$  & $|C_i|$ & $d$ & $H_i$ \\ \hline
 $1$   & $0$ & $1$ & $0$ & $I_2$ \\
 $2$     & $p^jI_2$ ($j=0,1$)& $p^{2(2-j)m}$ & $1$ & $p^{2-j}I_2$ \\
 $2$     & $(w p,p)$  & $p^{m}$ & $2$ & $\left[\begin{array}{cc}-w & 1 \cr 0 & p\end{array}\right]$ \\
  $2$     & $\left[\begin{array}{cc}w & 1 \cr 0 & p\end{array}\right]$ & $p^{3m}$ & $1$ & $(-wp,p)$ \\
  $2$     & $(w,1)$     & $p^{2m}$   &  $2$                         & $(-w,1)$
\\ \hline
\end{tabular}
\end{center}
\textit{where $w\in\{1,-1\}$, $d={\rm d}_{H}^{(A_i)}(C_i)$, $N_i$ is the number of $C_i$ in the same line and $I_2$ is the identity matrix of order $2$ (the same below)}.

\par
  (i-2) \textit{If $p=2$, $(G_i,H_i)$ is given by the following table}.

\begin{center}
\begin{tabular}{lcll|c}\hline
  $N_i$  &  $G_i$  & $|C_i|$ & $d$ & $H_i$ \\ \hline
$1$   & $0$ & $1$ & $0$ & $I_2$ \\
 $2$     & $2^jI_2$ ($j=0,1$) & $2^{2(2-j)m}$ & $1$ & $2^{2-j}I_2$ \\
 $1$     & $(2,2)$ & $2^{m}$ & $2$ & $\left[\begin{array}{cc}1 & 1 \cr 0 & 2\end{array}\right]$ \\
  $1$     & $\left[\begin{array}{cc}1 & 1 \cr 0 & 2\end{array}\right]$ & $2^{3m}$ & $1$ & $(2,2)$ \\
 $2^m$     & $(1+2u,1)$  ($u\in \mathcal{T}$)   & $2^{2m}$   &  $2$                         & $(-1+2u,1)$
\\ \hline
\end{tabular}
\end{center}
\textit{where $\mathcal{T}$ is a Teichm\"{u}ller set of $R$ with $|\mathcal{T}|=|\mathbb{F}_{2^m}|=2^m$ and $\overline{\mathcal{T}}=\mathbb{F}_{2^m}$}.

\par
  (ii) \textit{Let $d_i\geq 2$ and $\zeta_i(x)$ be an invertible element
of $A_i$ with multiplicative order $p^{md_i}-1$. Then $d_i$ is even and  $(G_i,H_i)$ is given by the following table}.

\begin{center}
\begin{tabular}{lcll|c}\hline
  $N_i$  &  $G_i$  & $|C_i|$ & $d$ & $H_i$ \\ \hline
 $1$   & $0$ & $1$ & $0$ & $I_2$ \\
 $2$     & $p^jI_2$ ($j=0,1$) & $p^{2(2-j)m}$ & $1$ & $p^{2-j}I_2$ \\
 $p^{\frac{md_i}{2}}+1$     & $(pw(x),p)$ & $p^{md_i}$ & $2$ & $\left[\begin{array}{cc}-w(x) & 1 \cr 0 & p\end{array}\right]$ \\
 $p^{\frac{md_i}{2}}+1$     & $\left[\begin{array}{cc}w(x) & 1 \cr 0 & p\end{array}\right]$ & $p^{3md_i}$ & $1$ & $(-pw(x),p)$ \\
 $p^{md_i}+p^{\frac{md_i}{2}}$     & $(w(x)(1+p\vartheta(x)),1)$   & $p^{2md_i}$   &  $2$                         & $(-w(x)(1+p\vartheta(x)),1)$
\\ \hline
\end{tabular}
\end{center}
\textit{where $w(x)\in\mathcal{W}_i$ and $\vartheta(x)\in \mathcal{V}_i$. Here}

\par
  $\bullet$ $\mathcal{W}_i=\{\zeta_i(x)^{(p^{\frac{md_i}{2}}-1)s}\mid s=0,1,\ldots,p^{\frac{md_i}{2}}\}$;

\par
  $\bullet$ \textit{$\mathcal{V}_i=\{0\}\cup\{\overline{\zeta}_i(x)^{(2^{\frac{md_i}{2}}+1)l}\mid l=0,1,\ldots,2^{\frac{md_i}{2}}-2\}$ when
$p=2$, and $\mathcal{V}_i=\{0\}\cup\{\overline{\zeta}_i(x)^{\frac{1}{2}(p^{\frac{md_i}{2}}+1)+(p^{\frac{md_i}{2}}+1)l}\mid l=0,1,\ldots,p^{\frac{md_i}{2}}-2\}$ when $p$ is odd}.

\par
  (II) \textit{Let $r+1\leq i\leq r+t$ and $\zeta_i(x)$ be an invertible element
of $\Upsilon_{i,1}$ with multiplicative order $p^{md_i}-1$. Find $\phi_i(x),\psi_i(x)\in R[x]$ satisfying
\begin{equation}
\phi_i(x)\rho_i^\ast(x)+\psi_i(x)\rho_i(x)=1,
\end{equation}
and set $\epsilon_{i,1}(x), \epsilon_{i,2}(x)\in A_i=R[x]/\langle f_i(x)\rangle=R[x]/\langle \rho_i(x)\rho_i^\ast(x)\rangle$ as follows}

\vskip 2mm\par
  $\bullet$ \textit{$\epsilon_{i,1}(x)\equiv \phi_i(x)\rho_i^\ast(x)$ $({\rm mod} \ f_i(x))$ and $\epsilon_{i,2}(x)\equiv \psi_i(x)\rho_i(x)$ $({\rm mod} \ f_i(x))$}.

\vskip 2mm\noindent
\textit{Then $(G_i,H_i)$ is given by the following table}:

{\footnotesize \begin{center}
\begin{tabular}{lcll|c}\hline
  $N_i$  &  $G_i$  & $|C_i|$ & $d$ & $H_i$ \\ \hline
$1$   & $0$ & $1$ & $0$ & $I_2$ \\
 $2$     & $p^jI_2$ ($j=0,1$) & $p^{2(2-j)m}$ & $1$ & $p^{2-j}I_2$ \\
 $1$ &  $pM_{i,1}$
 & $p^{2md_i}$ & $1$ & $\left[\begin{array}{c}M_{i,1} \cr pM_{i,2}\end{array}\right]$ \\
 $1$ &  $\left[\begin{array}{c}M_{i,1} \cr pM_{i,2}\end{array}\right]$
 & $p^{6md_i}$ & $1$ & $pM_{i,1}$ \\
$1$ &  $pM_{i,2}$
 & $p^{2md_i}$ & $1$ & $\left[\begin{array}{c}M_{i,2} \cr pM_{i,1}\end{array}\right]$ \\
$1$ &  $\left[\begin{array}{c}M_{i,2} \cr pM_{i,1}\end{array}\right]$
 & $p^{6md_i}$ & $1$ & $pM_{i,2}$ \\
 $p^{md_i}-1$     & $(pw(x),p)$ & $p^{2md_i}$ & $2$ & $\left[\begin{array}{cc}-w(x) & 1 \cr 0 & p\end{array}\right]$ \\
 $p^{md_i}-1$     & $\left[\begin{array}{cc}w(x) & 1 \cr 0 & p\end{array}\right]$ & $p^{6md_i}$ & $1$ & $(-pw(x),p)$ \\
 $p^{md_i}$ &  $\left[\begin{array}{cc}\epsilon_{i,1}(x) & pb_{i,1}(x) \cr pb_{i,1}(x^{-1}) & \epsilon_{i,2}(x)\end{array}\right]$
 & $p^{4md_i}$ & $1$ & $\left[\begin{array}{cc}\epsilon_{i,1}(x) & -pb_{i,1}(x) \cr -pb_{i,1}(x^{-1}) & \epsilon_{i,2}(x)\end{array}\right]$ \\
 $p^{2md_i}-p^{md_i}$     & $(w(x)+p\vartheta(x),1)$   & $p^{4md_i}$   &  $2$                         & $(-w(x)-p\vartheta(x),1)$ \\
 $p^{md_i}$ &  $\left[\begin{array}{cc}\epsilon_{i,2}(x) & pb_{i,2}(x) \cr pb_{i,2}(x^{-1}) & \epsilon_{i,1}(x)\end{array}\right]$
 & $p^{4md_i}$ & $1$ & $\left[\begin{array}{cc}\epsilon_{i,2}(x) & -pb_{i,2}(x) \cr -pb_{i,2}(x^{-1}) & \epsilon_{i,1}(x)\end{array}\right]$ \\
 \hline
\end{tabular}
\end{center}}

\noindent
\textit{where $M_{i,1}=\left[\begin{array}{cc}\epsilon_{i,1}(x) & 0 \cr 0 & \epsilon_{i,2}(x)\end{array}\right]$,
$M_{i,2}=\left[\begin{array}{cc}\epsilon_{i,2}(x) & 0 \cr 0 & \epsilon_{i,1}(x)\end{array}\right]$,
$b_{i,j}(x)\in K_{i,j}$ for $j=1,2$, $\vartheta(x)\in \mathcal{V}_i^{(w(x))}$ and $w(x)\in\mathcal{W}_i$. Here}

\vskip 2mm\par
  $\bullet$ $U_{i,1}=\{0\}\cup\{\epsilon_{i,1}(x)\zeta_i(x)^k\mid k=0,1,\ldots,p^{md_i}-2\}$ (mod $f_i(x)$);

\vskip 2mm\par
  $\bullet$ $K_{i,1}=\{0\}\cup\{\overline{\epsilon}_{i,1}(x)\overline{\zeta}_i(x)^k\mid k=0,1,\ldots,p^{md_i}-2\}=U_{i,1}$ (mod $p$);

\vskip 2mm\par
  $\bullet$ $K_{i,2}=\{0\}\cup\{\overline{\epsilon}_{i,2}(x)\overline{\zeta}_i(x^{n-1})^k\mid k=0,1,\ldots,p^{md_i}-2\}$ (mod $\overline{f}_i(x)$);

\vskip 2mm\par
  $\bullet$ $\mathcal{W}_i=\{u(x)+\frac{1}{u(x^{-1})}\mid 0\neq u(x)\in U_{i,1}\}$;

\vskip 2mm\par
  $\bullet$ \textit{$\mathcal{V}_i^{(w(x))}=\{v(x)-(\frac{1}{u(x^{-1})})^2v(x^{-1})\mid v(x)\in U_{i,1}\}$ $({\rm mod} \ p)$
for any $w(x)=u(x)+\frac{1}{u(x^{-1})}\in \mathcal{W}_i$ where $0\neq u(x)\in U_{i,1}$}.

\vskip 3mm\noindent
   {\bf Remark} In Theorem 2.5(II), we have that $v(x^{-1})=0$ if $v(x)=0$. Now, let
$u(x)=\epsilon_{i,1}(x)\zeta_i(x)^k$ where $0\leq k\leq p^{md_i}-2$. Then

\vskip 2mm\par
  $\bullet$ $u(x^{-1})=\epsilon_{i,2}(x)\zeta_i(x^{-1})^k=\epsilon_{i,2}(x)\zeta_i(x^{n-1})^k$ (mod $f_i(x)$).

\vskip 2mm\par
  $\bullet$ $\frac{1}{u(x^{-1})}=\epsilon_{i,2}(x)\zeta_i(x^{-1})^{p^{md_i}-1-k}=\epsilon_{i,2}(x)\zeta_i(x^{n-1})^{p^{md_i}-1-k}$ (mod $f_i(x)$).

\vskip 2mm\par
  $\bullet$ $b_{i,1}(x^{-1})=\overline{\epsilon}_{i,2}(x)\overline{\zeta}_i(x^{n-1})^k\in K_{i,2}$, if $b_{i,1}(x)=\overline{\epsilon}_{i,1}(x)\overline{\zeta}_i(x)^k\in K_{i,1}$ where $0\leq k\leq p^{md_i}-2$.

\vskip 2mm\par
  $\bullet$ $b_{i,2}(x^{-1})=\overline{\epsilon}_{i,1}(x)\overline{\zeta}_i(x)^k\in K_{i,1}$, if $b_{i,1}(x)=\overline{\epsilon}_{i,2}(x)\overline{\zeta}_i(x^{n-1})^k\in K_{i,2}$ where $0\leq k\leq p^{md_i}-2$.

\vskip 3mm\par
   Finally, by Theorem 2.5 we obtain a formula to
count the number of all left $D_{2n}$-codes over $R$.

\vskip 3mm\noindent
   {\bf Corollary 2.6} \textit{Denote $\lambda=|\{i\mid d_i=1, 0\leq i\leq r\}|$. Let $\mathcal{N}_{(n,p^2,p^{2m})}$ be the
number of left $D_{2n}$-codes over $R$. If $p$ is odd,}
$$\mathcal{N}_{(n,p^2,p^{2m})}=9^{\lambda}\prod_{d_i\geq 2, 1\leq i\leq r}\left(p^{md_i}+3p^{\frac{md_i}{2}}+5\right)
\prod_{i=r+1}^{r+t}\left(p^{2md_i}+3p^{md_i}+5\right);$$
\textit{and if $p=2$},
$$\mathcal{N}_{(n,4,4^{m})}=(2^m+5)^{\lambda}\prod_{d_i\geq 2, 1\leq i\leq r}\left(2^{md_i}+3\cdot 2^{\frac{md_i}{2}}+5\right)
\prod_{i=r+1}^{r+t}\left(4^{md_i}+3\cdot 2^{md_i}+5\right).$$

%%%%%%%%%%%%%%%%%%%%%%%%%%%%%%%%%%%%%%%%%%%%%%%%%%%%%%%%%%%%%%%%%%%%%%%%%%%%%%%%%%%%

%%%%%%%%%%%%%%%%%%%%%%%%%%%%%%%%%%%%%%%%%%%%%%%%%%%%%%%%%%%%%%%%%%%%%%%%%%%%%%%%%%%%
\section{Proof of Theorem 2.5}
\noindent
  In this section, we give a proof for Theorem 2.5. First, for basic irreducible (basic primitive)
polynomials in $R[x]$ we have the following conclusion.

\vskip 3mm\noindent
  {\bf Lemma 3.1} (cf. [36] Theorem 14.22) \textit{For any integer $\kappa\geq 1$ there exist monic basic irreducible
(and monic basic primitive) polynomials of degree $\kappa$ over $R$ and dividing $x^{p^{m\kappa}-1}-1$ in $R[x]$}.

\vskip 3mm
\par
   Since $n$ is a positive integer satisfying ${\rm gcd}(p,n)=1$, there is an integer $\kappa$ such that $\kappa={\rm min}\{s\in \mathbb{Z}^{+}\mid (p^m)^s\equiv 1 \ ({\rm mod} \ n)\}$. Then $n$ is a divisor of $p^{m\kappa}-1$. Using Lemma 3.1, we choose a fixed monic
basic irreducible polynomial $\varsigma(z)$ of degree $\kappa$ in $R[z]$, and set
$\widehat{R}=R[z]/\langle \varsigma(z)\rangle$ which is a Galois ring of characteristic $p^2$ and cardinality $p^{2m\kappa}$ (cf. [36] Theorem 14.23). Now, we choose an invertible element $\zeta$ of $\widehat{R}$ with multiplicative order $p^{m\kappa}-1$ (cf. [36] Theorem 14.27) and set
$\omega=\zeta^{\frac{p^{m\kappa}-1}{n}}$. Then ${\rm ord}(\omega)=n$ and
$$x^n-1=\prod_{i=0}^{n-1}(x-\omega^i).$$
  For each $0\leq i\leq n-1$, let

\par
  $\diamond$ $r_i={\rm min}\{l\in\mathbb{Z}^{+}\mid i(p^m)^l\equiv i \ ({\rm mod} \ n)\}$;

\par
  $\diamond$ $J_i^{(p^m)}=\{i,ip^m,ip^{2m},\ldots, ip^{(r_i-1)m}\}$.

\noindent
Then $J_i^{(p^m)}$ is the $p^m$-cyclotomic coset modulo $n$ containing $i$, and the minimal polynomial of $\omega^i$ over $R$ is given by
$$M_i(x)=\prod_{s\in J_i^{(p^m)}}(x-\omega^{s})=\prod_{l=0}^{r_i-1}(x-\omega^{ip^{lm}})$$
which is a monic basic irreducible polynomial in $R[x]$ and satisfies $M_i(x)|(x^n-1)$.
Moreover, we have
$$M_i^\ast(x)=x^{r_i}\prod_{l=0}^{r_i-1}(\frac{1}{x}-\omega^{ip^{lm}})=u\prod_{l=0}^{r_i-1}(x-\omega^{-ip^{lm}})
=uM_j(x),$$
where $0\leq j\leq n-1$ satisfying $j\equiv-i$ (mod $n$) and $u=\omega^{i(\sum_{l=0}^{r_i-1}p^{lm})}\in R^{\times}$ (cf. [17] Lemma 2.3(ii)).
Then we have the following conclusion.

\vskip 3mm \noindent
  {\bf Lemma 3.2} \textit{$M_i(x)$ is a self-reciprocal polynomial
if and only if $ J_i^{(p^m)}= J_{-i \ ({\rm mod} \ n)}^{(p^m)}$, i.e., $i(p^{lm}+1)\equiv 0$ $($mod $n)$ for some integer $l$, $0\leq l\leq n-1$}.

\vskip 3mm \par
   From now on, we define

\par
  $\diamond$ $\theta(a(x))=a(x^{-1})=a(x^{n-1}) \ ({\rm mod} \ x^n-1)$,  $\forall a(x)\in \mathcal{A}=R[x]/\langle x^n-1\rangle.$

\noindent
Then one can easily verify that
$\theta$ is a ring automorphism on ${\cal A}$ satisfying $\theta^2={\rm id}_{ \mathcal{A}}$ and  $\theta(r)=r$ for any $r\in R$.

\vskip 3mm \noindent
   {\bf Lemma 3.3} \textit{Using the notations in Lemma 2.3 and Section 2, in the ring ${\cal A}$ the following hold}.

\vskip 2mm\par
  (i) \textit{$\theta(\varepsilon_i(x))=\varepsilon_i(x)$, $\theta({\cal A}_i)={\cal A}_i$
and the restriction $\theta|_{{\cal A}_i}$ of $\theta$ on ${\cal A}_i$ is a ring automorphism on ${\cal A}_i$, for all
$0\leq i\leq r+t$}.

\vskip 2mm\par
  (ii) \textit{For each $0\leq i\leq r+t$, define $\theta_i: A_i\rightarrow A_i$ by
$\theta_i(b(x))=b(x^{n-1})$ $({\rm mod} \ f_i(x))$. Then the following diagram for
ring isomorphisms commutes}
$$\left.\begin{array}{ccc}\ \ \ \ A_i & \stackrel{\theta_i}{\longrightarrow} & A_i\cr
                          \varphi_i\downarrow & &\ \ \ \ \downarrow  \varphi_i\cr
                         \ \ \ \  {\cal A}_i & \stackrel{\theta |_{\mathcal{A}_i}}{\longrightarrow} &{\cal A}_i\end{array}\right.$$
\textit{i.e., $\theta_i=\varphi_i^{-1}(\theta|_{{\cal A}_i})\varphi_i$. Hence $\theta_i^2={\rm id}_{A_i}$}.

\vskip 3mm \noindent
   {\bf Proof.} (i) As $f_i(x)$ is self-reciprocal, by Lemma 3.2 and the assumption of $f_i(x)$ it follows that
$f_i(x)=\prod_{j\in J}(x-\omega^{j})$,
where
$J=J_{s}^{(p^m)}\cup J_{-s}^{(p^m)}$ and $J_{-s}^{(p^m)}=\{-j \ ({\rm mod} \ n)\mid j\in J_{s}^{(p^m)}\}$ for some
$0\leq s\leq n-1$. From this, by Equations (1) and the definition of $\varepsilon_i(x)$ in Section 2 we deduce that
$\varepsilon_i(\omega^{j})=1$ if $j\in J$ and $\varepsilon_i(\omega^{j})=0$ otherwise, $0\leq j\leq n-1$, which implies
$$\varepsilon_i(x)=\frac{1}{n}\sum_{k=0}^{n-1}(\sum_{j\in J}\omega^{-jk})x^k.$$
Then by $-J=J$ (mod $n$) and $x^n=1$ in $\mathcal{A}$, it follows that
\begin{eqnarray*}
\theta(\varepsilon_i(x))&=&\varepsilon_i(x^{-1})=\frac{1}{n}\sum_{k=0}^{n-1}(\sum_{j\in J}\omega^{-jk})x^{-k}=\frac{1}{n}\sum_{k=0}^{n-1}(\sum_{j\in J}\omega^{-(-j)(-k)})x^{-k}\\
  &=&\frac{1}{n}\sum_{k^\prime=0}^{n-1}(\sum_{j^\prime=-j, \ j\in J}\omega^{-j^\prime k^\prime})x^{k^\prime}=\varepsilon_i(x).
\end{eqnarray*}
Hence $\theta({\cal A}_i)=\theta(\varepsilon_i(x){\cal A})=\theta(\varepsilon_i(x))\theta({\cal A})=\varepsilon_i(x){\cal A}={\cal A}_i$.
Therefore, the restriction $\theta|_{{\cal A}_i}$ of $\theta$ on ${\cal A}_i$ is a ring automorphism of ${\cal A}_i$.

\par
  (ii) By the definition of $\varphi_i$ in Lemma 2.3(iii), we see that the inverse $\varphi_i^{-1}$ of $\varphi_i$
is a ring isomorphism from ${\cal A}_i$ onto $A_i$ satisfying
$$\varphi_i^{-1}(w(x))\equiv w(x) \ ({\rm mod} \ f_i(x)), \forall w(x)\in {\cal A}_i,$$
which implies $\varphi_i^{-1}(\varepsilon_i(x))\equiv 1-v_i(x)f_i(x)\equiv 1$ $({\rm mod} \ f_i(x))$, i.e.,
$\varphi_i^{-1}(\varepsilon_i(x))=1$ in $A_i$. Then for any $a(x)\in A_i$, by (i) and Lemma 2.3(iii) it follows that
\begin{eqnarray*}
\varphi_i^{-1}(\theta|_{{\cal A}_i})\varphi_i(a(x))&=&\varphi_i^{-1}(\theta|_{{\cal A}_i}(\varphi_ia(x)))=\varphi_i^{-1}(\theta|_{{\cal A}_i}(\varepsilon_i(x)a(x)))\\
  &=&\varphi_i^{-1}(\theta(\varepsilon_i(x))a(x^{-1}))=\varphi_i^{-1}(\varepsilon_i(x))\varphi_i^{-1}(a(x^{-1}))\\
  &=&a(x^{-1}) \ ({\rm mod} \ f_i(x))\\
  &=&\theta_i(a(x)).
\end{eqnarray*}
Hence $\theta_i=\varphi_i^{-1}(\theta|_{{\cal A}_i})\varphi_i$.
\hfill $\Box$

\vskip 3mm \par
   Now, we define the skew polynomial ring over $\mathcal{A}$ with indeterminate $y$ by

\par
 $\diamond$ ${\cal A}[y;\theta]=\{\sum_{j=0}^k\alpha_jy^j\mid
\alpha_j\in {\cal A}, \ j=0,1,\ldots,k, \ k=0,1,\ldots\}$,
where the multiplication is determined by
$$ya(x)=\theta(a(x))y=a(x^{-1})y, \forall a(x)\in {\cal A}.$$
Since $y^2a(x)=a(x)y^2$ for all $a(x)\in {\cal A}$, $y^2-1$ generates a two-sided ideal $\langle y^2-1\rangle$
of ${\cal A}[y;\theta]$. In this paper, we denote

\par
  $\diamond$ ${\cal R}={\cal A}[y;\theta]/\langle y^2-1\rangle=\{a(x)+b(x)y\mid a(x),b(x)\in {\cal A}\}$ $(y^2=1)$

\noindent
which is the residue class ring of ${\cal A}[y;\theta]$ modulo its ideal $\langle y^2-1\rangle$.

\par
   Similarly, for each $0\leq i\leq r+t$ by Lemma 3.3(ii) we can define the skew polynomial ring over $A_i$ by

\par
  $\diamond$ $A_i[y;\theta_i]=\{\sum_{j=0}^k a_j(x)y^j\mid
a_j(x)\in A_i, \ j=0,1,\ldots,k, \ k=0,1,\ldots\}$,
where the multiplication is determined by
$$yb(x)=\theta_i(b(x))y=b(x^{n-1})y \ ({\rm mod} \ f_i(x)), \forall b(x)\in A_i.$$
As $f_i(x)$ is a minic divisor of $x^n-1$ in $R[x]$, $x^{n-1}=x^{-1}$ in $A_i$ and $y^2-1$ generates a two-sided ideal $\langle y^2-1\rangle$
of $A_i[y;\theta_i]$. In this paper, we denote

\par
  $\diamond$ $R_i=A_i[y;\theta_i]/\langle y^2-1\rangle=\{b_0(x)+b_1(x)y\mid b_0(x),b_1(x)\in A_i\}$ $(y^2=1)$

\noindent
which is the residue class ring of $A_i[y;\theta_i]$ modulo its ideal $\langle y^2-1\rangle$.

\vskip 3mm \noindent
   {\bf Lemma 3.4} \textit{Let $0\leq i\leq r+t$. We extend the ring isomorphism $\varphi_i$ from
$A_i$ onto $\mathcal{A}_i\subseteq \mathcal{A}$ to the following map}
$$b_0(x)+b_1(x)y\mapsto \varphi_i(b_0(x))+\varphi_i(b_1(x))y=\varepsilon_i(x)(b_0(x)+b_1(x)y) \
({\rm mod} \ x^n-1)$$
\textit{$(\forall b_0(x)+b_1(x)y\in R_i)$, and also denote this map by $\varphi_i$. Then}

\par
   (i) \textit{The map $\varphi_i$ is an injective ring isomorphism from $R_i$ to $\mathcal{R}$}.

\vskip 2mm \par
   (ii) \textit{The mapping $\Phi: R_0\times R_1\times\ldots\times R_{r+t}\rightarrow {\cal R}$
defined by
$$\Phi(\beta_0,\beta_1,\ldots,\beta_{r+t})=\sum_{i=0}^{r+t}\varphi_i(\beta_i) \
(\forall \beta_i\in R_i, \ i=0,1,\ldots,r+t)$$
is an isomorphism of rings}.

\vskip 3mm \noindent
  {\bf Proof.} (i) For any $b(x)\in A_i$, by Lemma 3.3(ii) it follows that
\begin{eqnarray*}
\varphi_i(y b(x))&=&\varphi_i(\theta_i(b(x)y))=\varphi_i(\theta_i(b(x))y=(\varphi_i\theta_i)(b(x))y\\
&=&(\theta|_{{\cal A}_i}\varphi_i)(b(x))y
=\theta(\varphi_i(b(x)))y=y\varphi_i(b(x)).
\end{eqnarray*}
Since $\varphi_i$ is a ring isomorphism from $A_i$ onto $\mathcal{A}_i$ by Lemma 2.3(iii), we conclude that $\varphi_i$ is an injective homomorphism of rings from $R_i$ to $\mathcal{R}$.

\par
  (ii) By (i) and its proof, we conclude that $\Phi$ is a ring homomorphism from $R_0\times R_1\times\ldots\times R_{r+t}$ to $R$.
For any $\beta_i=b_{i0}(x)+b_{i1}(x)y\in R_i$ where $b_{i0}(x),b_{i1}(x)\in A_i$, it is clear that
$(\beta_0,\beta_1,\ldots,\beta_{r+t})\in {\rm Ker}(\Phi)$ if and only if $\sum_{i=0}^{r+t}\varphi_i(\beta_i)=0$ in $\mathcal{R}$,
where
\begin{eqnarray*}
\sum_{i=0}^{r+t}\varphi_i(\beta_i)&=&\sum_{i=0}^{r+t}\varepsilon_i(x)(b_{i0}(x)+b_{i1}(x)y) \ \ ({\rm by \ (i)}) \\
 &=&(\sum_{i=0}^{r+t}\varepsilon_i(x)b_{i0}(x))+(\sum_{i=0}^{r+t}\varepsilon_i(x)b_{i0}(x))y
\end{eqnarray*}
and $\varepsilon_i(x)b_{i0}(x),\varepsilon_i(x)b_{i1}(x)\in \mathcal{A}_i$ by Lemma 2.3(iii). As
$\mathcal{A}=\oplus_{i=0}^{r+t}\mathcal{A}_i$ by Lemma 2.3(ii), we deduce that
\begin{eqnarray*}
\sum_{i=0}^{r+t}\varphi_i(\beta_i)=0&\Longleftrightarrow &\sum_{i=0}^{r+t}\varepsilon_i(x)b_{i0}(x)=\sum_{i=0}^{r+t}\varepsilon_i(x)b_{i1}(x)=0
\ {\rm in} \ \mathcal{A} \\
&\Longleftrightarrow & \varepsilon_i(x)b_{i0}(x)=\varepsilon_i(x)b_{i1}(x)=0 \ (\forall i=0,1,\ldots,r+t),
\end{eqnarray*}
where $\varepsilon_i(x)b_{i0}(x)=\varphi_i(b_{i0}(x)), \varepsilon_i(x)b_{i1}(x)=\varphi_i(b_{i1}(x))\in \mathcal{A}_i$ for all
$i=0,1,\ldots,r+t$. Since $\varphi_i$ is a ring isomorphism from $A_i$ onto $\mathcal{A}_i$ by Lemma 2.3(iii), we conclude that
$(\beta_0,\beta_1,\ldots,\beta_{r+t})\in {\rm Ker}(\Phi)$ if and only if $b_{i0}(x)=b_{i1}(x)=0$, i.e. $\beta_i=0$, for all
$i=0,1,\ldots,r+t$. So ${\rm Ker}(\Phi)=\{(0,0,\ldots,0)\}$ and hence $\Phi$ is injective. Moreover, by Lemma 2.3 (iii) and (ii) we have
\begin{eqnarray*}
|R_0\times R_1\times\ldots\times R_{r+t}|&=&\prod_{i=0}^{r+t}|R_i|=\prod_{i=0}^{r+t}|A_i|^2=\prod_{i=0}^{r+t}|\mathcal{A}_i|^2
 =(\prod_{i=0}^{r+t}|\mathcal{A}_i|)^2\\
 &=&|\mathcal{A}|^2=|\mathcal{R}|.
\end{eqnarray*}
Therefore, $\varphi_i$ is a ring isomorphism from $R_i$ onto $\mathcal{R}$.
\hfill $\Box$

\vskip 3mm \par
  Let $0\leq i\leq r+t$ and $C_i$ be an $A_i$-submodule of $A_i^2=\{(b_0(x),b_1(x))\mid b_0(x),b_1(x)\in A_i\}$.
  Recall that $C_i$ is called a \textit{skew $\theta_i$-cyclic code} if
  $$(\theta_i(b_1(x)),\theta_i(b_0(x)))\in C_i, \ \forall (b_0(x),b_1(x))\in C_i.$$
 From now on, we will identify each $(b_0(x),b_1(x))\in C_i$ with $b_0(x)+b_1(x)y\in R_i=A_i[y;\theta_i]/\langle y^2-1\rangle$. Then it is clear that $C_i$ is a skew $\theta_i$-cyclic code over $A_i$ of length $2$
if and only if $C_i$ is a left ideal of $R_i$ (See [6, Theorem 1]).
For more details on skew cyclic codes, readers are referred to
[5]--[9] and [26].

\par
   Now, we give a direct sum decomposition for each left $D_{2n}$-code over $R$ by the following lemma.

\vskip 3mm \noindent
  {\bf Lemma 3.5} \textit{For any $\alpha=\sum_{i=0}^{n-1}a_{i,0}x^i+\sum_{i=0}^{n-1}a_{i,1}x^iy\in R[D_{2n}]$
where $a_{i,0},a_{i,1}$ $\in R$ for all $i=0,1,\ldots,n-1$, we identify $\alpha$ with
$a_0(x)+a_1(x)y\in \mathcal{R}$ where $a_0(x)=\sum_{i=0}^{n-1}a_{i,0}x^i,a_1(x)=\sum_{i=0}^{n-1}a_{i,1}x^i\in \mathcal{A}$. Then the following statements are equivalent}:

\vskip 2mm\par
  (i) \textit{$\mathcal{C}$ is a left $D_{2n}$-code over $R$};

\vskip 2mm\par
  (ii) \textit{$\mathcal{C}$ is a left ideal of the ring $\mathcal{R}=\mathcal{A}[y;\theta]/\langle y^2-1\rangle$};

\vskip 2mm\par
  (iii) \textit{For each $0\leq i\leq r+t$, there is a unique skew $\theta_i$-cyclic code $C_i$ over $A_i$ of length $2$ such
that
$${\cal C}=\oplus_{i=0}^{r+t}({\cal A}_i\Box_{\varphi_i}C_i).$$
Therefore, $|{\cal C}|=\prod_{i=0}^{r+t}|C_i|$}.

\vskip 3mm \noindent
  {\bf Proof.} (i)$\Leftrightarrow$(ii). As $y^2=1$
and $yx=x^{-1}y$ in the group $D_{2n}$, we see that the identification of $R[D_{2n}]$ with $\mathcal{R}$ is a ring isomorphism.
Hence $\mathcal{C}$ is a left ideal of $R[D_{2n}]$ if and only if $\mathcal{C}$ is a left ideal of $\mathcal{R}$.

\par
  (ii)$\Leftrightarrow$(iii). By Lemma 3.4(ii), we deduce that $\mathcal{C}$ is a left ideal of $\mathcal{R}$
if and only if for each $0\leq i\leq r+t$, there is a unique left ideal $C_i$ of $R_i$ such
that
\begin{eqnarray*}
\mathcal{C}&=&\Phi(C_0\times C_1\times\ldots\times C_{r+t})=\oplus_{i=0}^{r+t}\varphi_i(C_i)\\
 &=&\oplus_{i=0}^{r+t}\{\varphi_i(b_{i,0}(x))+\varphi_i(b_{i,0}(x))y\mid b_{i,0}(x)+b_{i,1}(x)y\in C_i\}\\
  &=&\oplus_{i=0}^{r+t}\{(\varphi_i(b_{i,0}(x)),\varphi_i(b_{i,0}(x))\mid (b_{i,0}(x),b_{i,1}(x))\in C_i\}\\
  &=&\oplus_{i=0}^{r+t}({\cal A}_i\Box_{\varphi_i}C_i).
\end{eqnarray*}
In this case, we have $|\mathcal{C}|=|C_0\times C_1\times\ldots\times C_{r+t}|=\prod_{i=0}^{r+t}|C_i|$.
\hfill $\Box$

\vskip 3mm\par
  By Lemma 3.5, in order to list all left $D_{2n}$-codes over $R$ it is sufficiency
to determine  all left ideals of the ring $R_i=A_i[y;\theta_i]/\langle y^2-1\rangle$
where $A_i=R[x]/\langle f_i(x)\rangle$ for all $i=0,1,\ldots,r+t$. Using the notations of Section 2, we have the following
two cases:

\vskip 3mm\par
  ({\bf I}) $0\leq i\leq r$.

\par
   In this case, $f_i(x)$ is a monic basic irreducible and self-reciprocal polynomial in $R[x]$.
By Lemma 2.2(i), $A_i=R[x]/\langle f_i(x)\rangle$ is a Galois ring
of characteristic $p^2$ and cardinality $p^{2md_i}$, and there exists

\par
 $\diamond$ $\zeta_i(x)=\sum_{j=0}^{d_i-1}\zeta_{ij}x^j\in A_i^\times$ with $\zeta_{ij}\in R$  for all $0\leq j\leq d_i-1$
such that
${\rm ord}(\zeta_i(x))=p^{md_i}-1$ in $A_i$.

\noindent
Hence each element $\beta$ of $A_i$ has a unique $p$-expansion:
\begin{equation}
\beta=b_0(x)+pb_1(x), \ b_0(x),b_1(x)\in \langle \zeta_i(x)\rangle\cup\{0\}
\end{equation}
where $\langle \zeta_i(x)\rangle\cup\{0\}$ is a Teichm\"{u}ller set of $A_i$ with $\langle \zeta_i(x)\rangle=\{\zeta_i(x)^k\mid k=0,1,\ldots,p^{md_i}-2\}\subset A_i^\times$ (cf. [36] Theorem 14.27).

\par
   By [36] Section 14.6 and Equation (3), the \textit{generalized Frobenius automorphism} $\phi_i$ of $A_i$ over $R$ is defined by
\begin{equation}
\phi_i(b_0(x)+pb_1(x))=b_0(x)^{p^m}+pb_1(x)^{p^m}, \ \forall  b_0(x),b_1(x)\in \langle \zeta_i(x)\rangle\cup\{0\}.
\end{equation}
The set of all automorphism of $A_i$ over $R$ form a group with respect to the map composition of maps, which is called the
\textit{Galois group} of $A_i$ over $R$ and is denoted by ${\rm Gal}(A_i/R)$. It is known that the multiplicative order
of $\phi_i$ is equal to $d_i$ and
\begin{equation}
{\rm Gal}(A_i/R)=\langle \phi_i\rangle=\{\phi_i^0,\phi_i,\phi_i^2,\ldots,\phi_i^{d_i-1}\}.
\end{equation}

\par
  As $^{-}$ is a ring homomorphism from $R$ onto $\mathbb{F}_{p^m}$ defined
at the beginning of Section 2,
it follows that $\overline{f}_i(x)$ is a monic irreducible and self-reciprocal polynomial in $\mathbb{F}_{p^m}[x]$
of degree $d_i$. From now on, we denote

\par
   $\diamond$ $K_i=\mathbb{F}_{p^m}[x]/\langle \overline{f}_i(x)\rangle$ and $\overline{\zeta}_i(x)=\sum_{j=0}^{d_i-1}\overline{\zeta}_{ij}x^j\in K_i$
with $\overline{\zeta}_{ij}\in \mathbb{F}_{p^m}$.

\noindent
   Then $K_i$ is an extension field of $\mathbb{F}_{p^m}$ with degree $d_i$, which implies
$|K_i|=p^{md_i}$.  By Equation (3), the surjective ring homomorphism $^{-}: R\rightarrow \mathbb{F}_{p^m}$ can be extended to a surjective  homomorphism of rings from $A_i$ onto $K_i$, which is also denoted by $^{-}$, as follows
$$\beta=b_0(x)+pb_1(x)\mapsto \overline{\beta}=\left\{\begin{array}{ll}\overline{\zeta}_i(x)^k \ ({\rm mod} \ \overline{f}_i(x)) & {\rm if} \ b_0(x)=\zeta_i(x)^k;
  \cr 0 & {\rm if} \ b_0(x)=0. \end{array}\right.$$
where $b_0(x),b_1(x)\in \langle \zeta_i(x)\rangle\cup\{0\}$. Hence $\overline{\zeta}_i(x)$ is a primitive element of $K_i$, i.e, ${\rm ord}(\overline{\zeta}_i(x))=p^{md_i}-1$. Therefore,
$$K_i=\{0\}\cup\{\overline{\zeta}_i(x)^k\mid k=0,1,\ldots, p^{md_i}-2\}.$$

\par
   By [36] Section 7.1, the \textit{Frobenius automorphism} $\sigma_i$ of $K_i$ over $\mathbb{F}_{p^m}$ is defined by
$$\sigma_i(\alpha)=\alpha^{p^m}, \ \forall \alpha\in K_i.$$
The set of all automorphism of $K_i$ over $\mathbb{F}_{p^m}$ form a group with respect to the map composition of maps, which is called the
\textit{Galois group} of $K_i$ over $\mathbb{F}_{p^m}$ and is denoted by ${\rm Gal}(K_i/\mathbb{F}_{p^m})$. It is known that the multiplicative order
of $\sigma_i$ is equal to $d_i$ and
${\rm Gal}(K_i/\mathbb{F}_{p^m})=\langle \sigma_i\rangle=\{\sigma_i^0,\sigma_i,\sigma_i^2,\ldots,\sigma_i^{d_i-1}\}.$

\vskip 3mm\noindent
  {\bf Lemma 3.6} \textit{Using the notations above, we have the following conclusions}.

\vskip 2mm\par
  (i) (cf. [36] Theorem 14.32(iii)) \textit{The following diagram commutes}:
$$\left.\begin{array}{ccc}A_i& \stackrel{\phi_i}{\longrightarrow} & A_i \cr
                           ^{-}\downarrow &   &\downarrow ^{-}\cr
                           K_i& \stackrel{\sigma_i}{\longrightarrow} & K_i
                           \end{array}\right..$$

\par
  (ii) \textit{Let $d_i>1$ where $1\leq i\leq r$. Then $d_i$ is even and
$\theta_i=\phi_i^{\frac{d_i}{2}}$. Hence}
$$\theta_i(\xi)=\xi^{p^{\frac{md_i}{2}}}, \ \forall \xi\in \langle \zeta_i(x)\rangle\cup\{0\}.$$

\par
  (iii) \textit{Let $d_i>1$, $1\leq i\leq r$, and $w(x)=\zeta_i(x)^{(p^{\frac{md_i}{2}}-1)s}$ where $0\leq s\leq p^{\frac{md_i}{2}}$. Then
$w(x)\cdot \theta_i(w(x))=1$, i.e., $\theta_i(w(x))=w(x)^{-1}$}.

\vskip 3mm \noindent
  {\bf Proof.} (ii) As $d_i>1$ and $1\leq i\leq r$, by Lemma 3.2 there exists a lest positive integer $j$ such that $i(p^m)^j\equiv -i$ (mod $n$), which implies $i(p^m)^{2j}\equiv -i(p^m)^j\equiv i$ (mod $n$), and so
$d_i={\rm min}\{l\in \mathbb{Z}^{+}\mid i(p^m)^l\equiv i \ ({\rm mod} \ n)\}$ being a divisor of $2j$. As
$i(p^m)^j\equiv -i$ (mod $n$), $d_i$ is not a  divisor of $j$. Hence $d_i=2j$.

\par
  As stated above, by Equation (5) we see that ${\rm Gal}(A_i/R)$ is a cyclic group generated by $\phi_i$
with even order $d_i$. Since $\theta_i$ is an automorphism of $A_i$ over $R$ of order $2$, we conclude that
$\theta_i=\phi_i^{\frac{d_i}{2}}$. From this and by Equation (4), we deduce that
$\theta_i(\xi)=\phi_i^{\frac{d_i}{2}}(\xi)=\xi^{(p^m)^{\frac{d_i}{2}}}=\xi^{p^{\frac{md_i}{2}}}$ for all
$\xi\in \langle \zeta_i(x)\rangle\cup\{0\}$.

\par
  (iii) It follows that
$w(x)\cdot \theta_i(w(x))=\zeta_i(x)^{((p^{\frac{md_i}{2}}-1)+p^{\frac{md_i}{2}}(p^{\frac{md_i}{2}}-1))s}=1$,
by (ii) and $\zeta_i(x)^{p^{md_i}-1}=1$.
\hfill $\Box$

\vskip 3mm \par
  In the following, we adopt the following notation.

\par
   $\diamond$ Let $\overline{\theta}_i: K_i\rightarrow K_i$ be defined by
$\overline{\theta}_i(c(x))=c(x^{n-1}) \ ({\rm mod} \ \overline{f}_i(x))$,
for all $c(x)=\sum_{j=0}^{d_i-1}c_jx^j\in K_i$ with $c_j\in \mathbb{F}_{p^m}.$

\noindent
Then $\overline{\theta}_i$ is a ring automorphism
of $K_i$ such that the following diagram commutes:
\begin{equation}
\left.\begin{array}{ccc}A_i& \stackrel{\theta_i}{\longrightarrow} & A_i \cr
                           ^{-}\downarrow &   &\downarrow ^{-}\cr
                           K_i& \stackrel{\overline{\theta}_i}{\longrightarrow} & K_i
                           \end{array}\right.,
\ {\rm i.e.}, \ \overline{\theta_i(\beta)}=\overline{\theta}_i(\overline{\beta}), \forall \beta\in A_i.
\end{equation}
Moreover, by Lemma 3.6(ii) we deduce that $\overline{\theta}_i=\sigma_i^{\frac{d_i}{2}}$ when $d_i>1$ and
$1\leq i\leq r$. Hence $\overline{\zeta}_i(x)^{(p^{\frac{md_i}{2}}-1)s}\cdot\overline{\theta}_i(\overline{\zeta}_i(x)^{(p^{\frac{md_i}{2}}-1)s})=1$
for any integer $0\leq s\leq p^{\frac{md_i}{2}}$ by Lemma 3.6(iii).

\par
  $\diamond$ Let $K_i[y;\overline{\theta}_i]$ be the skew polynomial ring over the finite
field $K_i$ determined by $\overline{\theta}_i$.

\par
   $\diamond$ Denote $\Gamma_i=K_i[y;\overline{\theta}_i]/\langle y^2-1\rangle$
which is the residue class ring of $K_i[y;\overline{\theta}_i]$ modulo its two-sided ideal $\langle y^2-1\rangle$
generated by $y^2-1$.

\noindent
  Then we extend the surjective ring homomorphism $^{-}: A_i\rightarrow K_i$ to a map from
$R_i$ onto $\Gamma_i$, which is denoted by $^{-}$ as well, by the natural way:
$$\overline{\beta_0+\beta_1y}=\overline{\beta}_0+\overline{\beta}_1y, \ \forall \beta_0,\beta_1\in  A_i.$$
For any $\beta\in A_i$, by Equation (6) it follows that
$$\overline{y\beta}=\overline{\theta_i(\beta)y}=\overline{\theta_i(\beta)}y=\overline{\theta}_i(\overline{\beta})y=y\overline{\beta}.$$
From this, it can be verify easily that $^{-}$ is a surjective ring homomorphism from $R_i$ onto $\Gamma_i=K_i[y;\overline{\theta}_i]/\langle y^2-1\rangle$.

\par
  By $\mathcal{W}_i=\{\zeta_i(x)^{(p^{\frac{md_i}{2}}-1)s}\mid 0\leq s\leq p^{\frac{md_i}{2}}\}\subseteq A_i^\times$ (see Section 2), we have

\par
  $\diamond$ $\overline{\mathcal{W}}_i=\{\overline{w}(x)\mid w(x)\in\mathcal{W}_i\}=\{\overline{\zeta_i}(x)^{(p^{\frac{md_i}{2}}-1)s}\mid 0\leq s\leq p^{\frac{md_i}{2}}\}
  \subseteq K_i^\times$.

\noindent
Then $\overline{\mathcal{W}}_i$ is the unique subgroup of $K_i^\times$ with order $p^{\frac{md_i}{2}}+1$.

\par
   For ideals of $\Gamma_i=K_i[y;\overline{\theta}_i]/\langle y^2-1\rangle$, we known the following conclusion.

\vskip 3mm \noindent
   {\bf Lemma 3.7} \textit{Let $d_i>1$. We have the following conclusions}:

\vskip 2mm\par
   (i) (cf. [18, Theorem 3.3]) \textit{Using the notations above, all distinct left
ideals of the ring $\Gamma_i=\mathbb{F}_{p^m}[y;\overline{\theta}_i]/\langle y^2-1\rangle$ are given by the following}:
$$\langle 0\rangle, \ \Gamma_i, \ \Gamma_i(\overline{w}(x)+y) \
 {\rm where} \ w(x)\in\mathcal{W}_i.$$
   \textit{Therefore, the number of left ideals of $\Gamma_i$ is equal to $p^{\frac{md_i}{2}}+3$}.

\vskip 2mm\par
   (ii) (cf. [18, Theorem 3.4 and Corollary 3.5]) \textit{For any $w(x)\in\mathcal{W}_i$,
$\Gamma_i(\overline{w}(x)+y)$ is a minimal left ideal of $\Gamma_i$,
$|\Gamma_i(\overline{w}(x)+y)|=p^{md_i}$, the minimum Hamming weight ${\rm w}_H^{(K_i)}(\Gamma_i(\overline{w}(x)+y))$
of $\Gamma_i(\overline{\zeta}_i(x)^{(p^{\frac{md_i}{2}}-1)s}+y)$ over $K_i$ is equal to $2$ and
$\Gamma_i=\Gamma_i(\overline{w}_1(x)+y)\oplus\Gamma_i(\overline{w}_2(x)+y)$ for any $w_1(x),w_2(x)\in\mathcal{W}_i$
satisfying $w_1(x)\neq w_2(x)$}.

\vskip 3mm\par
  Using the notations above, we give the following conclusions for left ideals of the ring $R_i=A_i[y;\theta_i]/\langle y^2-1\rangle$.

\vskip 3mm\noindent
  {\bf Theorem 3.8} \textit{Let $0\leq i\leq r$ and ${\rm deg}(f_i(x))=d_i=1$. Then $R_i=R[y]/\langle y^2-1\rangle$ which is a commutative ring.}

\vskip 2mm \par
  (i) (cf. [32] Definition 4.1 and Theorem 4.4)\textit{If $p$ is odd, there are $9$ distinct ideals of $R_i$ which are given by}:
$\{0\}$, $R_ip$, $R_ip(y-1)$, $R_ip(y+1)$, $R_i$, $R_i(y-1)$, $R_i(y+1)$, $R_i(y-1)+R_ip$,  $R_i(y+1)+R_ip$.

\par
  \textit{Moreover, we have $|\{0\}|=1$, $|R_i|=p^{4m}$, $|R_ip(y-1)|=|R_ip(y+1)|=p^m$, $|R_i(y-1)|=|R_i(y+1)|=p^{2m}$,
$|R_i(y-1)+R_ip|=|R_i(y+1)+R_ip|=p^{3m}$}.

\vskip 2mm \par
  (ii) (cf. [28] Theorem 3.8) \textit{If $p=2$, there are $2^m+5$ distinct ideals of $R_i$ which are given by}:
$\{0\}, \ 2R_i, \ 2R_i(y-1), \ R_i, \ R_i(y-1)+2R_i, \ R_i((y-1)+2u)\ {\rm with} \ u\in \mathcal{T},$
\textit{where $\mathcal{T}$ is a Teichm\"{u}ller set of $R$}.

\par
   \textit{Moreover, we have $|\{0\}|=1$, $|R_i|=2^{4m}$, $|2R_i|=|R_i((y-1)+2u)|=2^{2m}$, $|2R_i(y-1)|=2^m$
and $|R_i(y-1)+2R_i|=2^{3m}$}.

\vskip 3mm\noindent
  {\bf Theorem 3.9} \textit{Let $0\leq i\leq r$ and ${\rm deg}(f_i(x))=d_i\geq 2$. Then all skew $\theta_i$-cyclic code $C_i$ over $A_i$ of length $2$, i.e., all left ideals
of the ring $R_i=A_i[y;\theta_i]/\langle y^2-1\rangle$, are given by the following table}:
\begin{center}
\begin{tabular}{lllll}\hline
case &  $N_i$  &  $C_i$ (left ideals of $R_i$) & $|C_i|$ & $d$ \\ \hline
(1)  & $1$  & $\diamond$ $\{0\}$ & $0$  & $0$ \\
(2)  & $2$  & $\diamond$  $R_ip^j$ \ ($j=0,1$) & $p^{2(2-j)md_i}$  & $1$ \\
(3)  & $p^{\frac{md_i}{2}}+1$  & $\diamond$  $R_ip(w(x)+y)$ \ ($w(x)\in \mathcal{W}_i$) & $p^{md_i}$  & $2$ \\
(4)  & $p^{md_i}+p^{\frac{md_i}{2}}$  & $\diamond$  $R_i(w(x)(1+p\vartheta(x))+y)$ & $p^{2md_i}$  & $2$ \\
     &   & \ \ \  ($w(x)\in \mathcal{W}_i$, $\vartheta(x)\in \mathcal{V}_i$) &   & \\
(5)  & $p^{\frac{md_i}{2}}+1$  & $\diamond$  $R_i(w(x)+y)+R_ip$  \ ($w(x)\in \mathcal{W}_i$) & $p^{3md_i}$  & $1$ \\
\hline
\end{tabular}
\end{center}

\noindent
 \textit{where $N_i$ is the number of ideals in the same line of the table, $d={\rm w}_{H}^{(A_i)}(C_i)$ is the minimum Hamming weight of $C_i$ as a linear code over $A_i$ of length $2$}.

\par
  \textit{Therefore, the number of left ideals in $R_i$ is equal to $p^{md_i}+3p^{\frac{md_i}{2}}+5$}.

\vskip 3mm\noindent
  {\bf Proof.}  For any nonzero left ideal $C$ of $R_i$, let $\overline{C}=\{\overline{\alpha}\mid \alpha\in C\}$. Then
it is clear that $\overline{C}$ is a left ideal of $\Gamma_i$ and $\tau_i:\alpha\mapsto \overline{\alpha}$
($\forall\alpha\in C$) is an $A_i$-linear homomorphism from $C$ onto $\overline{C}$. Hence ${\rm Im}(\tau_i)
=\overline{C}$. Let $(C:p)=\{\alpha \in R_i\mid p\alpha\in C\}$. It is known that
$(C:p)$ is a left ideal of $R_i$ satisfying $C\subseteq (C:p)$, which implies that $\overline{(C:p)}$
is a left ideal of $\Gamma_i$  satisfying $\overline{C}\subseteq \overline{(C:p)}$. Moreover, by
${\rm Ker}(\tau_i)=\{\alpha\in C\mid \overline{\alpha}=0\}=p\overline{(C:p)}$ it follows that
\begin{equation}
|C|=|{\rm Im}(\tau_i)||{\rm Ker}(\tau_i)|=|\overline{C}||\overline{(C:p)}|.
\end{equation}
Then by Lemma 3.7, we have one of the following cases.

\par
  (i) $\overline{C}=\{0\}$. As $C\neq \{0\}$, by (7) it follows that $\overline{(C:p)}$ is a
nonzero left ideal of $\Gamma_i$. Then by Lemma 3.7 we have one of the following subcases:

\par
  (i-1) $\overline{(C:p)}=\Gamma_i$. Then $C\subseteq R_ip$ and $1\in \overline{(C:p)}$. The latter implies
$1+p\alpha\in (C:p)$ for some $\alpha\in R_i$. Then by $p^2=0$ and the definition of
$(C:p)$, we have $p=p(1+p\alpha)\in C$, which implies $R_ip\subseteq C$. Hence $C=R_ip$.

\par
  By Lemma 3.7(ii) and Equation (7), we deduce that
$|C|=|\Gamma_i|=p^{2md_i}$ and ${\rm w}_{H}^{(A_i)}(C)={\rm w}_{H}^{(K_i)}(\overline{(C:p)})
={\rm w}_{H}^{(K_i)}(\Gamma_i)=1$.

\par
  (i-2) $\overline{(C:p)}=\Gamma_i(\overline{w}(x)+y)$ where $w(x)\in \mathcal{W}_i$.
Then $\overline{w(x)+y}=\overline{w}(x)+y\in \overline{(C:p)}$, which implies
$w(x)+y+p\alpha\in (C:p)$ for some $\alpha\in R_i$, and so $p(w(x)+y)=p(w(x)+y+p\alpha)\in C$.
Hence $R_ip(w(x)+y)\subseteq C$.

\par
  Conversely, let $\xi\in C$. By $\overline{C}=\{0\}$, there exists $\beta\in R_i$ such that
$\xi=p\beta$, which implies $\beta\in (C:p)$, and so $\overline{\beta}\in \overline{(C:p)}$.
Then there exists $\gamma,\delta\in R_i$ such that $\beta=\gamma(w(x)+y+p\alpha)+p\delta$, which implies
$$\xi=p(\gamma(w(x)+y+p\alpha)+p\delta)=\gamma\cdot p(w(x)+y)\in R_ip(w(x)+y).$$
Hence $C\subseteq R_ip(w(x)+y)$ and so $C=R_ip(w(x)+y)$.

\par
  Then by Lemma 3.7(ii) and Equation (7), we have that
$|C|=|\Gamma_i(\overline{w}(x)+y)|=p^{md_i}$ and ${\rm w}_{H}^{(A_i)}(C)={\rm w}_{H}^{(K_i)}(\overline{(C:p)})
={\rm w}_{H}^{(K_i)}(\Gamma_i(\overline{w}(x)+y))=2$.

\par
  (ii) $\overline{C}=\Gamma_i$. Then $1+p\alpha\in C$ for some $\alpha\in R_i$. As $C\subseteq (C:p)$, we
have $\overline{(C:p)}=\Gamma_i$. From this and by the proof of (i-2), we know that $p\in C$. Hence
$1=(1+1+p\alpha)-\alpha\cdot p\in C$. Therefore $C=R_i=R_ip^0$ and hence $|C|=|R_i|=p^{4md_i}$. It is obvious that
${\rm w}_{H}^{(A_i)}(C)=1$ in this case.

\par
  (iii) $\overline{C}=\Gamma_i(\overline{w}(x)+y)$ where $w(x)\in \mathcal{W}_i$.
By $\overline{(y+w(x))}=\overline{w}(x)+y\in C$, there exists $\alpha\in R_i$ such that
$y+w(x)+p\alpha\in C$.  As $\overline{C}\subseteq \overline{(C:p)}$, by Lemma 3.8(ii) we
have one of the following two subcases:

\par
  (iii-1) $\overline{(C:p)}=\Gamma_i$. In this case, by the proof of (i-1), we know that $p\in C$. Hence
$y+w(x)=(y+w(x)+p\alpha)-\alpha\cdot p\in C$, and so $R_i(y+w(x))+R_ip\subseteq C$.

\par
  Conversely, let $\xi\in C$. By $\overline{\xi}\in \overline{C}=\Gamma_i(y+\overline{w}(x))$, there
exists $\gamma,\delta\in R_i$ such that $\xi=\gamma(y+w(x))+p\delta\in R_i(y+w(x))+R_ip$.
Hence $C\subseteq R_i(y+w(x))+R_ip\subseteq C$ and so $C=R_i(y+w(x))+R_ip$.

\par
  Moreover, by Lemma 3.7(ii) and Equation (7) it follows that
$|C|=|\Gamma_i(\overline{w}(x)+y)||\Gamma_i|=p^{3md_i}$ and ${\rm w}_{H}^{(A_i)}(C)={\rm w}_{H}^{(K_i)}(\overline{(C:p)})
={\rm w}_{H}^{(K_i)}(\Gamma_i)=1$.

\par
  (iii-2) $\overline{(C:p)}=\Gamma_i((\overline{w}(x)+y)$. Then by the proof of (i-2), we know
that $p(\overline{w}(x)+y)=p(y+w(x))\in C$. As $\alpha\in R_i$, there exist $a(x),b(x)\in A_i$ such
that $\alpha=a(x)+b(x)y$. Denote
$$\tau(x)\equiv a(x)-b(x)w(x) \ ({\rm mod} \ p), \
{\rm where} \ \tau(x)\in \overline{A}_i=K_i.$$
Then $p\tau(x)=p(a(x)-b(x)w(x))$ and hence
\begin{eqnarray*}
y+w(x)+p\tau(x)&=&y+w(x)+p(a(x)+b(x)y)-p(b(x)y+b(x)w(x))\\
&=&(y+w(x)+p\alpha)-b(x)\cdot p(y+w(x))\in C.
\end{eqnarray*}
Therefore, $R_i(y+w(x)+p\tau(x))\subseteq C$.

\par
  Conversely, let $\xi\in C$. By $\overline{C}=\Gamma_i(\overline{w}(x)+y)=\overline{R_i(y+w(x)+p\tau(x))}$, there
exist $\gamma,\delta\in R_i$ such that
$\xi=\gamma(y+w(x)+p\tau(x))+p\delta$, which implies $p\delta=\xi-\gamma(y+w(x)+p\tau(x))\in C$,
i.e., $\delta\in (C:p)$. From this and by  $\overline{(C:p)}=\Gamma_i((\overline{w}(x)+y)$, we deduce
that $\delta=\rho(y+w(x))+p\eta$ for some $\rho,\eta\in R_i$. Therefore,
\begin{eqnarray*}
\xi&=&\gamma(y+w(x)+p\tau(x))+p(\rho(y+w(x))+p\eta)\\
&=&\gamma(y+w(x)+p\tau(x))+\rho\cdot p((y+w(x)+p\tau(x))\\
&=&(\gamma+p\rho)(y+w(x)+p\tau(x))\in R_i(y+w(x)+p\tau(x)).
\end{eqnarray*}
Hence $C=R_i(y+w(x)+p\tau(x))$.

\par
   Suppose that $C=R_i(y+w(x)+p\nu(x))$ for some $\nu(x)\in K_i$
as well. Then $p(\tau(x)-\nu(x))=(y+w(x)+p\tau(x))-(y+w(x)+p\nu(x))\in C$,
which implies $\tau(x)-\nu(x)\in \overline{(C:p)}=\Gamma_i(\overline{w}(x)+y)$ with
$\vartheta(x)-\nu(x)\in K_i$. Since ${\rm w}_{H}^{(K_i)}(\Gamma_i(\overline{w}(x)+y))=2$ by Lemma 3.7(ii),
we have $\tau(x)-\nu(x)=0$, i.e., $\tau(x)=\nu(x)$ in $K_i$.

\par
  By Lemma 3.6(iii), we see that $w(x)\cdot \theta_i(w(x))=1$, i.e., $\theta_i(w(x))=w(x)^{-1}$.
From this and by $p^2=0$, we deduce that
$\left(\theta_i(w(x))+p\theta_i(\tau(x))\right)^{-1}=w(x)^{2}\left(\theta_i(w(x))-p\theta_i(\tau(x))\right).$
As $C$ is a left ideal of $R_i$ and $y^2=1$, we have
\begin{eqnarray*}
&& y+w(x)+p(-\overline{w}(x)^{2}\theta_i(\tau(x)))\\
&=&y+w(x)^{2}\left(\theta_i(w(x))-p\theta_i(\tau(x))\right)\\
&=&\left(\theta_i(w(x))+p\theta_i(\tau(x))\right)^{-1}\left(\left(\theta_i(w(x))+p\theta_i\tau(x))\right)y+1\right)\\
&=&\left(\theta_i(w(x))+p\theta_i(\tau(x))\right)^{-1}y\cdot(w(x)+p\vartheta(x)+y)\in C,
\end{eqnarray*}
which implies $-\overline{w}(x)^{2}\theta_i(\tau(x))=\tau(x)$ in $K_i$, i.e.,
$$\overline{\theta}_i\left(\overline{\theta}_i(\overline{w}(x))\tau(x)\right)+\overline{\theta}_i(\overline{w}(x))\tau(x)
=\overline{w}(x)\overline{\theta}_i(\tau(x))+\overline{\theta}_i(\overline{w}(x))\tau(x)=0$$
since $\overline{\theta}_i(\overline{w}(x))=\overline{w}(x)^{-1}$. Denote $\vartheta(x)=\overline{\theta}_i(\overline{w}(x))\tau(x)\in K_i$. Then
$\tau(x)=\overline{w}(x)\vartheta(x)$ and $p\tau(x)=pw(x)\vartheta(x)$ where $\vartheta(x)$ satisfies the following equation
\begin{equation}
\vartheta(x)^{p^{\frac{md_i}{2}}}+\vartheta(x)=\overline{\theta}_i(\vartheta(x))+\vartheta(x)=0,
\end{equation}
as $\overline{\theta}_i(\xi)=\sigma_i^{\frac{d_i}{2}}(\xi)=\xi^{p^{\frac{md_i}{2}}}$ for all $\xi\in K_i$.

\par
  $\diamondsuit$ Let $p=2$. Then Equation (8) is equivalent to $(\vartheta(x)^{2^{\frac{md_i}{2}}-1}-1)\vartheta(x)=0$.
Since $\overline{\zeta}_i(x)$ is a primitive
element of the finite field $K_i$ with multiplicative order $2^{md_i}-1$, in this case Equation (8) has exactly $2^{\frac{md_i}{2}}$ solutions
in $K_i$: $\vartheta(x)\in \mathcal{V}_i$ where
$\mathcal{V}_i=\{0\}\cup\{\overline{\zeta}_i(x)^{(2^{\frac{md_i}{2}}+1)l}\mid l=0,1,\ldots,2^{\frac{md_i}{2}}-2\}.$

\par
  $\diamondsuit$ Let $p$ be an odd prime.
Since $\overline{\zeta}_i(x)$ is a primitive
element of the finite field $K_i$ with multiplicative order $p^{md_i}-1$, we have $\overline{\zeta}_i(x)^{\frac{p^{md_i}-1}{2}}=-1$.
Then Equation (8) is equivalent to $\vartheta(x)=0$ or
\begin{equation}
\vartheta(x)^{p^{\frac{md_i}{2}}-1}-\overline{\zeta}_i(x)^{\frac{p^{md_i}-1}{2}}=\vartheta(x)^{p^{\frac{md_i}{2}}-1}+1=0.
\end{equation}
Let $\vartheta(x)=\overline{\zeta}_i(x)^k$ where $0\leq k\leq p^{md_i}-2$. Then Equation (9) is equivalent to
$k(p^{\frac{md_i}{2}}-1)\equiv \frac{p^{md_i}-1}{2}$ (mod $p^{md_i}-1$), i.e.,
$$k\equiv \frac{1}{2}(p^{\frac{md_i}{2}}+1) \ ({\rm mod} \ p^{\frac{md_i}{2}}+1).$$
Therefore, Equation (8) has exactly $p^{\frac{md_i}{2}}$ solutions
in $K_i$: $\vartheta(x)\in \mathcal{V}_i$ where
$\mathcal{V}_i=\{0\}\cup\{\overline{\zeta}_i(x)^{\frac{1}{2}(p^{\frac{md_i}{2}}+1)+(p^{\frac{md_i}{2}}+1)l}\mid l=0,1,\ldots,p^{\frac{md_i}{2}}-2\}.$

\par
  As stated above, we conclude that $C=R_i(y+w(x)+p\tau(x))=R_i(y+w(x)(1+p\vartheta(x)))$ is uniquely determined by the pair $(w(x),\vartheta(x))$ of
polynomials where $w(x)\in \mathcal{W}_i$ and $\vartheta(x)\in \mathcal{V}_i$. Hence
the number of left ideals is equal to $(p^{\frac{md_i}{2}}+1)p^{\frac{md_i}{2}}=p^{md_i}+p^{\frac{md_i}{2}}$ in this case.

\par
   Then by Lemma 3.7(ii) and Equation (7), we have
$|C|=|\Gamma_i(\overline{\zeta}_i(x)^k+y)|^2=p^{2md_i}$ and ${\rm w}_{H}^{(A_i)}(C)={\rm w}_{H}^{(K_i)}(\overline{(C:p)})
={\rm w}_{H}^{(K_i)}(\Gamma_i(\overline{\zeta}_i(x)^k+y))=2$.
\hfill $\Box$

\vskip 3mm\par
  ({\bf II}) $r+1\leq i\leq t$.

\par
  In this case, $f_{i}(x)=\rho_i(x)\rho_i^\ast(x)$ where $\rho_i(x)$ is a monic basic irreducible polynomial such that
$\rho_i(x)$ and $\rho_i^\ast(x)$ are coprime in $R[x]$. Using the notations in Section 2, we have $A_i=R[x]/\langle f_i(x)\rangle$ which is an extension commutative ring of $R$ and $|A_i|=|R|^{2d_i}=(p^{2m})^{2d_i}=p^{4md_i}$.

\par
  As $^{-}$ is a ring homomorphism from $R$ onto $\mathbb{F}_{p^m}$ defined
in Section 2, we see that $\overline{\rho}_i(x)$ is a mionic irreducible polynomial in $\mathbb{F}_{p^m}[x]$ of degree $d_i$,
the reciprocal polynomial of $\overline{\rho}_i(x)$ in $\mathbb{F}_{p^m}[x]$ is given by
$\overline{\rho}_i^\ast(x)=\overline{\rho_i^\ast(x)}$. Then $\overline{\rho}_i^\ast(x)$ is also a mionic irreducible polynomial in $\mathbb{F}_{p^m}[x]$ of degree $d_i$ satisfying ${\rm gcd}(\overline{\rho}_i(x),\overline{\rho}_i^\ast(x))=1$. Hence $\overline{f}_i(x)=\overline{\rho}_i(x)\overline{\rho}_i^\ast(x)$ which is a monic self-reciprocal polynomial in $\mathbb{F}_{p^m}[x]$
of degree $2d_i$. In the following, let

\vskip 2mm\par
  $\diamond$ $K_i=\mathbb{F}_{p^m}[x]/\langle \overline{f}_i(x)\rangle=\mathbb{F}_{p^m}[x]/\langle \overline{\rho}_i(x)\overline{\rho}_i^\ast(x)\rangle$.

\vskip 2mm\noindent
 Then we extend the surjective homomorphism of rings $^{-}:R\rightarrow \mathbb{F}_{p^m}$ to a surjective homomorphism of rings
from $A_i$ onto $K_i$, which is also denoted by $^{-}$, in the natural way:
$$^{-}:\sum_{j=0}^{2d_i-1}\beta_jx^j\mapsto \sum_{j=0}^{2d_i-1}\overline{\beta}_jx^j \ (\forall \beta_j\in A_i, \ j=0,1,\ldots,2d_i-1).$$

\par
  By Equation (2) in Section 2 and the definitions of $\epsilon_{i,1}(x)$ and $\epsilon_{i,2}(x)$ in Theorem 2.5(II),
from classical ring theory we
deduce the following lemma.

\vskip 3mm \noindent
  {\bf Lemma 3.10} (cf. [37] Theorem 2.7 and its proof)\textit{Using the notations above, we have the following conclusions}:

\vskip 2mm\par
  (i) \textit{$\epsilon_{i,1}(x)^2=\epsilon_{i,1}(x)$, $\epsilon_{i,2}(x)^2=\epsilon_{i,2}(x)$,
$\epsilon_{i,1}(x)\epsilon_{i,2}(x)=0$ and $\epsilon_{i,1}(x)+\epsilon_{i,2}(x)=1$ in $A_i$}.

\vskip 2mm\par
  (ii) \textit{$A_i=A_{i,1}\oplus A_{i,2}$, where $A_{i,j}=A_i\epsilon_{i,j}(x)$ for $j=1,2$}.

\vskip 2mm \par
  (iii) \textit{The map
$\chi: \Upsilon_{i,1}\times \Upsilon_{i,2}\rightarrow A_i$ defined by}
$$\chi(g_1(x),g_2(x))=\epsilon_{i,1}(x)g_1(x)+\epsilon_{i,2}(x)g_2(x) \ ({\rm mod} \ f_i(x)), \ \forall
g_j(x)\in \Upsilon_{i,j}$$
\textit{is a ring isomorphism where $\Upsilon_{i,1}=R[x]/\langle \rho_i(x)\rangle$ and $\Upsilon_{i,2}=R[x]/\langle \rho_i^\ast(x)\rangle$}.

\vskip 2mm \par
  (iv) \textit{$\theta_i(\epsilon_{i,1}(x))=\epsilon_{i,2}(x)$ and $\theta_i(\epsilon_{i,2}(x))=\epsilon_{i,1}(x)$. Then $\theta_i$
induces a ring isomorphism between $A_{i,1}$ and $A_{i,2}$. Hence $\theta_i$ is a ring automorphism of $A_i$ with multiplicative order $2$}.

\vskip 3mm \par
   By Lemma 2.2(ii), $\Upsilon_{i,1}$ is a Galois ring of characteristic $p^2$ and cardinality $p^{2md_i}$ and there exists
$$\zeta_i(x)=\sum_{k=0}^{d_i-1}\zeta_{ik}x^k\in \Upsilon_{i,1}^\times \ {\rm with} \
\zeta_{ik}\in R \ {\rm for} \ {\rm all} \ k$$
such that
${\rm ord}(\zeta_i(x))=p^{md_i}-1$. Then every element of $\Upsilon_{i,1}$ has a unique $p$-expansion:
$a_0(x)+pa_1(x)$ where $a_0(x),a_1(x)\in \{0\}\cup\{\zeta_i(x)^l\mid l=0,1,\ldots,p^{md_i}-1\}$ (mod $\rho_i(x)$). Let

\par
   $\diamond$ $\overline{\theta}_i: K_i\rightarrow K_i$ be defined by
$$\overline{\theta}_i(c(x))=c(x^{n-1}) \ ({\rm mod} \ \overline{f}_i(x)), \
\forall c(x)=\sum_{j=0}^{d_i-1}c_jx^j\in K_i \ {\rm with} \ c_j\in \mathbb{F}_{p^m}.$$
As $\overline{f}_i(x)\mid (x^n-1)$ in $\mathbb{F}_{p^m}[x]$, $\overline{\theta}_i$ is a ring automorphism
of $K_i$ such that the following diagram commutes:
\begin{equation}
\left.\begin{array}{ccc}A_i& \stackrel{\theta_i}{\longrightarrow} & A_i \cr
                           ^{-}\downarrow &   &\downarrow ^{-}\cr
                           K_i& \stackrel{\overline{\theta}_i}{\longrightarrow} & K_i
                           \end{array}\right.,
\ {\rm i.e.}, \ \overline{\theta_i(\beta)}=\overline{\theta}_i(\overline{\beta}), \forall \beta\in A_i.
\end{equation}

\vskip 3mm \noindent
  {\bf Lemma 3.11} \textit{Using the notations above, denote $\pi_{i,j}=\overline{\epsilon}_{i,j}(x)\in K_i$
and $K_{i,j}=K_i\pi_{i,j}$ for $j=1,2$. Then we have the following conclusions}:

\vskip 2mm\par
  (i) \textit{$\pi_{i,1}^2=\pi_{i,1}$, $\pi_{i,2}^2=\pi_{i,2}$, $\pi_{i,1}\pi_{i,2}=0$ and
$\pi_{i,1}\pi_{i,2}=1$ in $K_i$}.

\vskip 2mm\par
  (ii) \textit{$K_i=K_{i,1}\oplus K_{i,2}$ where $K_{i,j}=\overline{A}_{i,j}=K_i\pi_{i,j}$ for $j=1,2$}.

\vskip 2mm\par
  (iii) \textit{$K_{i,1}=\{0\}\cup\{\overline{\epsilon}_{i,1}(x)\overline{\zeta}_i(x)^k\mid k=0,1,\ldots,p^{md_i}-2\}$ and}
$K_{i,2}=\{0\}\cup\{\overline{\epsilon}_{i,2}(x)\overline{\zeta}_i(x^{-1})^k \mid k=0,1,\ldots,p^{md_i}-2\}$,
\textit{which are both finite fields of cardinality $p^{md_i}$}.

\par
  (iv) \textit{$\overline{\theta}_i(\pi_{i,1})=\pi_{i,2}$ and $\overline{\theta}_i(\pi_{i,2})=\pi_{i,1}$. Then $\overline{\theta}_i$
induces a field isomorphism between $K_{i,1}$ and $K_{i,2}$. Hence $\overline{\theta}_i$ is a ring automorphism of $K_i$ with multiplicative order $2$}.

\vskip 3mm \noindent
  {\bf Proof.} (i), (ii) and (iv) follow from Lemma 3.10 (i), (ii) and (iv), respectively.

\par
  (iii) As $\zeta_i(x)\in \Upsilon_{i,1}^\times$ and ${\rm ord}(\zeta_i(x))=p^{md_i}-1$ by Lemma 2.2(ii), every element of $\Upsilon_{i,1}$ has a unique $p$-expansion:
$b_0(x)+pb_1(x)$, $b_0(x), b_1(x)\in \mathcal{T}_{i,1}$ where $\mathcal{T}_{i,1}=\{0\}\cup\{\zeta_i(x)^k\mid k=0,1,\ldots,p^{md_i}-2\}$.
From this, by Lemma 3.10 (ii) and (iii), we deduce that $A_{i,1}=\{\epsilon_{i,1}(x)b_0(x)+p\epsilon_{i,1}(x)b_1(x)\mid b_0(x), b_1(x)$ $\in \mathcal{T}_{i,1}\}$ (mod $f_i(x)$), which implies
\begin{eqnarray*}
K_{i,1}&=&\overline{A}_{i,1}=\{\overline{\epsilon}_{i,1}(x)\overline{b}_0(x)\mid b_0(x)\in \mathcal{T}_{i,1}\}\\
 &=&\{0\}\cup\{\overline{\epsilon}_{i,1}(x)\overline{\zeta}_i(x)^k\mid 0\leq k\leq p^{md_i}-2\}.
\end{eqnarray*}
Hence $K_{i,2}=\overline{\theta}_i(K_{i,1})=\overline{\theta_i(A_{i,1})}=\{0\}\cup\{\overline{\epsilon}_{i,2}(x)\overline{\zeta}_i(x^{-1})^k \ \mid k=0,1,\ldots,p^{md_i}-2\}$ by Lemma 3.10(iv).
\hfill $\Box$

\vskip 3mm
 Now, let $K_i[y;\overline{\theta}_i]$ be the skew polynomial ring over the finite
commutative ring $K_i$ determined by $\overline{\theta}_i$ and denote

\par
   $\diamond$ $\Gamma_i=K_i[y;\overline{\theta}_i]/\langle y^2-1\rangle$
which is the residue class ring of $K_i[y;\overline{\theta}_i]$ modulo its two-sided ideal $\langle y^2-1\rangle$
generated by $y^2-1$.

\noindent
  Then we extend the surjective ring homomorphism $^{-}: A_i\rightarrow K_i$ to the a map from
$R_i$ onto $\Gamma_i$, denoted by $^{-}$ as well, by the natural way:
$$\overline{\beta_0+\beta_1y}=\overline{\beta}_0+\overline{\beta}_1y, \ \forall \beta_0,\beta_1\in  A_i.$$
For any $\beta\in A_i$, by Equation (10) it follows that
$$\overline{y\beta}=\overline{\theta_i(\beta)y}=\overline{\theta_i(\beta)}y=\overline{\theta}_i(\overline{\beta})y=y\overline{\beta}.$$
From this, it can be verify easily that $^{-}$ is a surjective ring homomorphism from $R_i$ onto $\Gamma_i$.

\par

\vskip 2mm\par
  As $U_{i,1}=\{0\}\cup\{\epsilon_{i,1}(x)\zeta_i(x)^k\mid k=0,1,\ldots,p^{md_i}-2\}$
and $\mathcal{W}_i=\{u(x)+\frac{1}{u(x^{-1})}\mid 0\neq u(x)\in U_{i,1}\}$ (see Theorem 2.5(II)), we have that

\par
  $\diamond$ $\overline{\mathcal{W}}_i=\{\overline{u}(x)+\frac{1}{\overline{u}(x^{-1})}\mid \overline{u}(x)=\overline{\epsilon}_{i,1}(x)\overline{\zeta_i}(x)^k, 0\leq k\leq p^{md_i}-2\}\subseteq K_{i}^\times$.
Obviously, $|\mathcal{W}_i|=|\overline{\mathcal{W}}_i|=p^{md_i}-1$.

\par
   For ideals of $\Gamma_i=K_i[y;\overline{\theta}_i]/\langle y^2-1\rangle$, we have the following conclusions.

\vskip 3mm \noindent
   {\bf Lemma 3.12} \textit{Let $r+1\leq i\leq t$. We have the following conclusions}:

\vskip 2mm \par
  (i) (cf. [18] Theorem 4.2) \textit{all distinct left ideals
of $\Gamma_i=K_i[y;\overline{\theta}_i]/\langle y^2-1\rangle$ are given by}:

\vskip 2mm\par
   \textit{$\{0\}$, $\Gamma_i$, $\Gamma_i\overline{\epsilon}_{i,1}$, $\Gamma_i\overline{\epsilon}_{i,2}$, $\Gamma_i(\overline{w}(x)+y)$
where $w(x)\in \mathcal{W}_i$}.

\vskip 2mm\noindent
  \textit{Moreover, we have the following properties for nonzero left
ideals of $\Gamma_i$}:

\vskip 2mm \par
  (i-1) \textit{$|\Gamma_i\overline{\epsilon}_{i,j}(x)|=p^{2md_i}$ and ${\rm wt}_H^{(K_i)}(\Gamma_i\overline{\epsilon}_{i,j}(x))=1$ for $j=1,2$;
$|\Gamma_i|=p^{4md_i}$ and ${\rm wt}_H^{(K_i)}(\Gamma_i)=1$}.

\vskip 2mm \par
  (i-2) \textit{$|J|=p^{2md_i}$ and ${\rm wt}_H^{(K_i)}(J)=2$ for any
$J=\Gamma_i(\overline{w}(x)+y)$ with $w(x)\in \mathcal{W}_i$}.

\vskip 2mm \par
   (ii) \textit{Let $\mathcal{L}=\{\Gamma_i\overline{\epsilon}_{i,j}(x)\mid j=1,2\}\cup\{\Gamma_i(\overline{w}(x)+y)\mid w(x)\in \mathcal{W}_i\}$. Then for any $J_1, J_2\in \mathcal{L}$ satisfying $J_1\neq j_2$, we have $\Gamma_i=J_1\oplus J_2$. Hence every
left ideal of $\Gamma_i$ contained in $\mathcal{L}$ is a minimal left ideal of $\Gamma_i$.}

\vskip 3mm\noindent
  {\bf Proof.} (ii) Let $J_1, J_2\in \mathcal{L}$ satisfying $J_1\neq J_2$. Then $|J_1|=|J_2|=p^{2md_i}$ by (i).
If $J_1\cap J_2=\{0\}$, we have $|J_1+J_2|=|J_1||J_2|=p^{4md_i}=|\Gamma_i|$, which implies $J_1+J_2=\Gamma_i$,
and hence $\Gamma_i=J_1\oplus J_2$. Suppose that $J_1\cap J_2\neq\{0\}$. Since $J_1\cap J_2$ is also a left
ideal of $\Gamma_i$ satisfying $J_1\cap J_2\subseteq J_s$ for all $s=1,2$, by (i) it follows that $|J_1\cap J_2|=p^{2md_i}$,
which implies $J_1=J_2$, and we get a contradiction. Therefore, $\Gamma_i=J_1\oplus J_2$.
\hfill $\Box$

\vskip 3mm \par
   Now, we list all left ideals
of the ring $R_i=A_i[y;\theta_i]/\langle y^2-1\rangle$ as follows.

\vskip 3mm\noindent
  {\bf Theorem 3.13} \textit{Let $r+1\leq i\leq t$. Using the notations above, all skew $\theta_i$-cyclic code $C_i$ over $A_i$ of length $2$, i.e., all left ideals
of the ring $R_i=A_i[y;\theta_i]/\langle y^2-1\rangle$, are given by the following table}:
\begin{center}
\begin{tabular}{lllll}\hline
case &  $N_i$  &  $C_i$ (left ideals of $R_i$) & $|C_i|$ & $d$ \\ \hline
(1)  & $1$  & $\diamond$  $\{0\}$ & $0$  & $0$ \\
(2)  & $2$  & $\diamond$  $R_ip^j$ \ ($j=0,1$) & $p^{4(2-j)md_i}$  & $1$ \\
(3)  & $2$  & $\diamond$  $R_ip\epsilon_{i,j}(x)$ \ ($j=1,2$) & $p^{2md_i}$  & $1$ \\
(4)  & $p^{md_i}-1$  & $\diamond$  $R_ip(w(x)+y)$ \ ($w(x)\in \mathcal{W}_i$)& $p^{2md_i}$  & $2$ \\
(5)  & $2$  & $\diamond$  $R_i\epsilon_{i,j}(x)+R_ip$ \ ($j=1,2$) & $p^{6md_i}$  & $1$ \\
(6)  & $2p^{md_i}$  & $\diamond$  $R_i(\epsilon_{i,j}(x)+pb_{i,j}(x)y)$ & $p^{4md_i}$  & $1$ \\
     &              &  \ \ \ ($b_{i,j}\in K_{i,j}$, $j=1,2$) & & \\
(7)  & $p^{md_i}-1$  & $\diamond$  $R_i(w(x)+y)+R_ip$ \ ($w(x)\in \mathcal{W}_i$) & $p^{6md_i}$  & $1$ \\
(8)  & $p^{2md_i}-p^{md_i}$  & $\diamond$  $R_i(w(x)+p\vartheta(x)+y)$ & $p^{4md_i}$  & $2$ \\
     &   & \ \ \ $(\vartheta(x)\in \mathcal{V}_i^{(w(x))}$, $w(x)\in \mathcal{W}_i$) \  &   &  \\
\hline
\end{tabular}
\end{center}

\noindent
 \textit{where $N_i$ is the number of left ideals in the same line of the table, $d={\rm w}_{H}^{(A_i)}(C_i)$ is the minimum Hamming weight of $C_i$ as a linear code over $A_i$}.

\par
   \textit{Therefore, the number of left ideals in $R_i$ is equal to $p^{2md_i}+3p^{md_i}+5$}.

\vskip 3mm\noindent
  {\bf Proof.}  For any nonzero left ideal $C$ of $R_i$, let $\overline{C}=\{\overline{\alpha}\mid \alpha\in C\}$ and $(C:p)=\{\alpha \in R_i\mid p\alpha\in C\}$. Similar to the proof of Theorem 3.9, it can be verified that
$(C:p)$ is a left ideal of $R_i$ satisfying $C\subseteq (C:p)$, which implies that both $\overline{C}$ and $\overline{(C:p)}$
are left ideals of $\Gamma_i$  satisfying $\overline{C}\subseteq \overline{(C:p)}$ and
\begin{equation}
|C|=|{\rm Im}(\tau_i)||{\rm Ker}(\tau_i)|=|\overline{C}||\overline{(C:p)}|.
\end{equation}
Then by Lemma 3.12(i), we have one of the following cases.

\par
  (i) $\overline{C}=\{0\}$. As $C\neq \{0\}$, by (11) it follows that $\overline{(C:p)}$ is a
nonzero left ideal of $\Gamma_i$. Then by Lemma 3.12(ii) we have one of the following subcases:

\par
  (i-1) $\overline{(C:p)}=\Gamma_i$. By Lemma 3.12 and an argument similar to (i-1) in the proof of Theorem 3.9,
it follows that $C=R_ip$, $|C|=|\Gamma_i|=p^{2md_i}$ and ${\rm w}_{H}^{(A_i)}(C)=1$.

\par
  (i-2) $\overline{(C:p)}=\Gamma_i\overline{\epsilon}_{i,j}(x)$, where $j=1,2$. Then $\overline{(C:p)}=\overline{R_i\epsilon_{i,j}(x)}$, which
implies that $\epsilon_{i,j}(x)+p\alpha\in (C:p)$ for some $\alpha\in R_i$, and hence $p\epsilon_{i,j}(x)=p(\epsilon_{i,j}(x)+p\alpha)\in C$. From this we
deduce that $R_ip\epsilon_{i,j}(x)\subseteq C$.

\par
  Conversely, let $\xi\in C$. By $\overline{C}=\{0\}$ there exists $\beta\in R_i$
such that $\xi=p\beta\in C$, which implies $\beta\in (C:p)$, and so $\overline{\beta}\in \overline{(C:p)}=\overline{R_i\epsilon_{i,j}(x)}$.
Then there exist $\gamma,\delta\in R_i$ such that $\beta=\gamma\epsilon_{i,j}(x)+p\delta$,
and hence $\xi=p(\gamma\epsilon_{i,j}(x)+p\delta)=\gamma(p\epsilon_{i,j}(x))\in R_ip\epsilon_{i,j}(x)$. Therefore,
$C\subseteq R_ip\epsilon_{i,j}(x)$ and so $C=R_ip\epsilon_{i,j}(x)$.

\par
  Moreover, by Lemma 3.12(i) and Equation (11) it following that $|C|=|\Gamma_i\overline{\epsilon}_{i,j}|=p^{2md_i}$
and ${\rm w}_{H}^{(A_i)}(C)={\rm w}_{H}^{(K_i)}(\overline{(C:p)})
={\rm w}_{H}^{(K_i)}(\Gamma_i\overline{\epsilon}_{i,j}(x))=1$.

\par
  (i-3) $\overline{(C:p)}=\Gamma_i(\overline{w}(x)+y)$ where $w(x)\in\mathcal{W}_i$. By Lemma 3.12 and an argument similar to (i-2) in the proof of Theorem 3.9,
it follows that $C=R_ip(w(x)+y)$, $|C|=|\Gamma_i(\overline{w}(x)+y)|=p^{2md_i}$ and ${\rm w}_{H}^{(A_i)}(C)=2$.

\par
  (ii) $\overline{C}=\Gamma_i$. By an argument similar to (ii) in the proof of Theorem 3.9,
it follows that $C=R_i=R_ip^0$, $|C|=|R_i|=p^{8md_i}$ and ${\rm w}_{H}^{(A_i)}(C)=1$.

\par
  (iii) $\overline{C}=\Gamma_i\overline{\epsilon}_{i,j}(x)$ where $j=1,2$. By $\overline{\epsilon}_{i,j}(x)\in \overline{C}$,
there exists $\alpha\in R_i$ such that $\epsilon_{i,j}(x)+p\alpha\in C$. As $\overline{C}\subseteq \overline{(C:p)}$, by Lemma 3.7(ii) we
have one of the following two subcases:

\par
  (iii-1) $\overline{(C:p)}=\Gamma_i$. In this case, by the proof of (i-1), we know that $p\in C$. Hence
$\epsilon_{i,j}(x)=(\epsilon_{i,j}(x)+p\alpha)-\alpha\cdot p\in C$. Therefore, $R_i\epsilon_{i,j}(x)+R_ip\subseteq C$.

\par
  Conversely, let $\xi\in C$. By $\overline{\xi}\in \overline{C}=\Gamma_i\overline{\epsilon}_{i,j}(x)=\overline{R_i\epsilon_{i,j}(x)}$, there
exists $\gamma,\delta\in R_i$ such that $\xi=\gamma\epsilon_{i,j}(x)+p\delta\in R_i\epsilon_{i,j}(x)+R_ip$.
Hence $C\subseteq R_i\epsilon_{i,j}(x)+R_ip\subseteq C$ and so $C=R_i\epsilon_{i,j}(x)+R_ip$.

\par
  Moreover, by Lemma 3.12(i) and Equation (11) it follows that
$|C|=|\Gamma_i\overline{\epsilon}_{i,j}||\Gamma_i|=p^{6md_i}$ and ${\rm w}_{H}^{(A_i)}(C)={\rm w}_{H}^{(K_i)}(\overline{(C:p)})
={\rm w}_{H}^{(K_i)}(\Gamma_i)=1$.

\par
  (iii-2) $\overline{(C:p)}=\Gamma_i\overline{\epsilon}_{i,j}(x)$. Then by the proof of (i-2), we know
that $p\overline{\epsilon}_{i,j}(x)=p\epsilon_{i,j}(x)\in C$, which implies $p\theta_i(\epsilon_{i,j}(x))y=y\cdot p\epsilon_{i,j}(x)\in C$.

\par
   $\diamondsuit$ Let $i=1$. In this case, we have that $p\epsilon_{i,1}(x)\in C$ and
$p\epsilon_{i,2}(x)y=p\theta_i(\epsilon_{i,1}(x))y\in C$ by Lemma 3.10(iv).
   As $\alpha\in R_i=A_i[y;\theta_i]/\langle y^2-1\rangle$, by Lemma 3.10(ii) there
exist $a_{i,1}(x),b_{i,1}(x)\in A_{i,1}$ and $a_{i,2}(x),b_{i,2}(x)\in A_{i,2}$ such that
$\alpha=(a_{i,1}(x)+a_{i,2}(x))+(b_{i,1}(x)+b_{i,2}(x))y$. By Lemma 3.10(i) and (ii), we have $a_{i,1}(x)\epsilon_{i,1}(x)=a_{i,1}(x)$,
$a_{i,2}(x)\epsilon_{i,1}(x)=0$, $b_{i,1}(x)y\epsilon_{i,1}(x)=b_{i,1}(x)\epsilon_{i,2}(x)y=0$ and $b_{i,2}(x)y\epsilon_{i,1}(x)=b_{i,2}(x)\epsilon_{i,2}(x)y=b_{i,2}(x)y$, which imply
$\alpha\cdot p\epsilon_{i,1}(x)=p(a_{i,1}(x)+b_{i,2}(x)y)$. Hence
$$\delta=\epsilon_{i,1}(x)+p(a_{i,2}(x)+b_{i,1}(x)y)=\epsilon_{i,1}(x)+p\alpha-\alpha\cdot p\epsilon_{i,1}(x)\in C.$$
Then by $\epsilon_{i,1}(x)^2=\epsilon_{i,1}(x)$, $\epsilon_{i,1}(x)a_{i,2}(x)=0$ and $\epsilon_{i,1}(x)b_{i,1}(x)=b_{i,1}(x)$, it follows that
$\epsilon_{i,1}(x)+pb_{i,1}(x)y=\epsilon_{i,1}(x)\delta\in C$, which implies $R_i(\epsilon_{i,1}(x)+pb_{i,1}(x)y)\subseteq C$.

\par
  Conversely, let $\xi\in C$. By $\overline{C}=\overline{R_i\epsilon_{i,1}(x)}=\Gamma_i\overline{\epsilon_{i,1}(x)+pb_{i,1}(x)y}$, there
exist $\gamma,\delta\in R_i$ such that
$\xi=\gamma(\epsilon_{i,1}(x)+pb_{i,1}(x)y)+p\delta$, which implies $p\delta=\xi-\gamma(\epsilon_{i,1}(x)+pb_{i,1}(x)y)\in C$,
i.e., $\delta\in (C:p)$. From this and by  $\overline{(C:p)}=\Gamma_i\overline{\epsilon}_{i,1}(x)$, we deduce
that $\delta=\lambda\epsilon_{i,1}(x)+p\eta$ for some $\lambda,\eta\in R_i$. Therefore,
$\xi=\gamma(\epsilon_{i,1}(x)+pb_{i,1}(x)y)+p(\lambda\epsilon_{i,1}(x)+p\eta)=(\gamma+p\lambda)(\epsilon_{i,1}(x)+pb_{i,1}(x)y)\in R_i(\epsilon_{i,1}(x)+pb_{i,1}(x)y)$.
Hence $C=R_i(\epsilon_{i,1}(x)+pb_{i,1}(x)y)$.

\par
  Assume that $C=R_i(\epsilon_{i,1}(x)+ph_{i,1}(x)y)$ from some $h_{i,1}\in K_{i,1}$ as well.
Then $p(b_{i,1}(x)-h_{i,1}(x))y=(\epsilon_{i,1}(x)+pb_{i,1}(x)y)-(\epsilon_{i,1}(x)+ph_{i,1}(x)y)\in C$, which implies
$(b_{i,1}(x)-h_{i,1}(x))y\in \overline{(C:p)}=\Gamma_i\overline{\epsilon}_{i,1}(x)$, and so
$$\overline{\theta}_i(b_{i,1}(x)-h_{i,1}(x))=y\cdot(b_{i,1}(x)-h_{i,1}(x))y\in \Gamma_i\overline{\epsilon}_{i,1}(x)=K_{i,1}+K_{i,2}y.$$
From this we deduce that $\overline{\theta}_i(b_{i,1}(x)-h_{i,1}(x))=0$, since $\overline{\theta}_i(b_{i,1}(x)-h_{i,1}(x))\in K_{i,2}$ and $K_{i,1}\cap K_{i,2}=\{0\}$ by Lemma 3.11 (iv) and (ii) respectively. Hence
$b_{i,1}(x)-h_{i,1}(x)=0$, i.e., $b_{i,1}(x)=h_{i,1}(x)$ in $K_{i,1}$.

\par
   As stated above, we conclude that all distinct left ideals of $R_i$ satisfying $\overline{C}=\overline{(C:p)}=\Gamma_i\overline{\epsilon}_{i,1}(x)$ are given by:
$C=R_i(\epsilon_{i,1}(x)+pb_{i,1}(x)y)$, where $b_{i,1}\in K_{i,1}$ and $|K_{i,1}|=p^{md_i}$.

\par
  $\diamondsuit$ Let $j=2$. An similar argument shows that  all distinct left ideals of $R_i$ satisfying $\overline{C}=\overline{(C:p)}=\Gamma_i\overline{\epsilon}_{i,2}(x)$ are given by:
$C=R_i(\epsilon_{i,2}(x)+pb_{i,2}(x)y)$, where $b_{i,2}\in K_{i,2}$ and $|K_{i,2}|=p^{md_i}$.

\par
  Moreover, by Lemma 3.12(i) and Equation (11) it follows that
$|C|=|\Gamma_i\overline{\epsilon}_{i,j}(x)|^2=p^{4md_i}$ and ${\rm w}_{H}^{(A_i)}(C)={\rm w}_{H}^{(K_i)}(\overline{(C:p)})
={\rm w}_{H}^{(K_i)}(\Gamma_i\overline{\epsilon}_{i,j}(x))=1$.

\par
  (iv) $\overline{C}=\Gamma_i(\overline{w}(x)+y)$ where $w(x)\in \mathcal{W}_i$.
By $\overline{y+w(x)}=\overline{w}(x)+y\in \overline{C}$, there exists $\alpha\in R_i$ such that
$y+w(x)+p\alpha\in C$.  As $\overline{C}\subseteq \overline{(C:p)}$, by Lemma 3.12(ii) we
have one of the following two subcases:

\par
  (iv-1) $\overline{(C:p)}=\Gamma_i$. By Lemma 3.12 and an argument similar to (iii-1) in the proof of Theorem 3.9,
it follows that $C=R_i(y+w(x))+R_ip$, $|C|=|\Gamma_i(\overline{w}(x)+y)||\Gamma_i|=p^{6md_i}$ and ${\rm w}_{H}^{(A_i)}(C)={\rm w}_{H}^{(K_i)}(\overline{(C:p)})
={\rm w}_{H}^{(K_i)}(\Gamma_i)=1$.

\par
  (iv-2) $\overline{(C:p)}=\Gamma_i(\overline{w}(x)+y)$. Then by the proof of (i-3), we know
that $p(y+w(x))\in C$. As $\alpha\in R_i$, there exist $a(x),b(x)\in A_i$ such
that $\alpha=a(x)+b(x)y$. Denote
$\vartheta(x)\equiv a(x)-b(x)w(x)$ $({\rm mod} \ p)$
where $\vartheta(x)\in \overline{A}_i=K_i$.
Then $p\vartheta(x)=p(a(x)-b(x)w(x))$ and hence
$$y+w(x)+p\vartheta(x)=(y+w(x)+p\alpha)-b(x)\cdot p(y+w(x))\in C$$
Hence $R_i(y+w(x)+p\vartheta(x))\subseteq C$. From this and by an argument similar to (iii-2) in the proof of Theorem 3.9,
we deduce that $C=R_i(y+w(x)+p\vartheta(x))$ and $C$ is uniquely determined by
the pair $(w(x),\vartheta(x))$ of polynomials.

\par
  \par
   By the definition of $\mathcal{W}_i$ (see Theorem 2.5(II)), we have that $w(x)=u(x)+\frac{1}{u(x^{-1})}=u(x)+(\theta_i(u(x)))^{-1}$
where $u(x)=\epsilon_{i,1}(x)\zeta_i(x)^k$ and $0\leq k\leq p^{md_i}-2$. By Lemma 3.10(iv), it follows
that $\theta_i(u(x))=\epsilon_{i,2}(x)\zeta_i(x^{-1})^k$ (mod $f_i(x)$), which implies
$\theta_i(w(x))=\theta_i(u(x))+(u(x))^{-1}$ where $\theta_i(u(x))\in A_{i,2}$ and $(u(x))^{-1}\in A_{i,1}$ by Lemma 3.10(iv). From this and by Lemma 3.10 (i) and (ii), we deduce that
$$w(x)\cdot \theta_i(w(x))=u(x)(u(x))^{-1}+(\theta_i(u(x)))^{-1}\theta_i(u(x))=\epsilon_{i,1}(x)+\epsilon_{i,2}(x)=1,$$
i.e., $\theta_i(w(x))=w(x)^{-1}$.
Then by $p^2=0$, we have
$\left(\theta_i(w(x))+p\theta_i(\vartheta(x))\right)^{-1}=w(x)^{2}\left(\theta_i(w(x))-p\theta_i(\vartheta(x))\right).$
As $C$ is a left ideal of $R_i$ and $y^2=1$, we have
\begin{eqnarray*}
&& y+w(x)+p(-\overline{w}(x)^{2}\overline{\theta}_i(\vartheta(x)))\\
&=&y+w(x)^{2}\left(\theta_i(w(x))-p\theta_i(\vartheta(x))\right)\\
&=&\left(\theta_i(w(x))+p\theta_i(\vartheta(x))\right)^{-1}\left(\left(\theta_i(w(x))+p\theta_i\vartheta(x))\right)y+1\right)\\
&=&\left(\theta_i(w(x))+p\theta_i(\vartheta(x))\right)^{-1}y\cdot(w(x)+p\vartheta(x)+y)\in C,
\end{eqnarray*}
From this, we deduce that
\begin{equation}
\overline{w}(x)\overline{\theta}_i(\vartheta(x))+\overline{\theta}_i(\overline{w}(x))\vartheta(x)=0, \ {\rm i.e.}, \
\vartheta(x)=-\overline{w}(x)^{2}\overline{\theta}_i(\vartheta(x))
\end{equation}
in $K_i$. Let $\overline{w}(x)=u(x)+\frac{1}{u(x^{-1})}$ where $u(x)=\overline{\epsilon}_{i,1}(x)\overline{\zeta}_i(x)^s\in K_{i,1}^\times$, $0\leq s\leq p^{md_i}-2$. By Equation (10), Lemma 3.11 and $u(x^{-1})=\overline{\theta}_i(u(x))\in K_{i,2}$,
it follows that $\overline{w}(x)^{2}=u(x)^2+\frac{1}{u(x^{-1})^2}$. By Lemma 3.11(ii), there exists a unique pair $(v_1(x),v_2(x))$,
$v_j(x)\in K_{i,j}$ for $j=1,2$, such that $\vartheta(x)=v_1(x)+v_2(x)$. Then
$\overline{\theta}_i(\vartheta(x))=\overline{\theta}_i(v_2(x))+\overline{\theta}_i(v_1(x))$ where
$\overline{\theta}_i(v_2(x))\in K_{i,1}$ and $\overline{\theta}_i(v_1(x))\in K_{i,2}$ by Lemma 3.11(iv).
Hence
$$\overline{w}(x)^{2}\overline{\theta}_i(\vartheta(x))=u(x)^2\overline{\theta}_i(v_2(x))+\frac{1}{u(x^{-1})^2}\overline{\theta}_i(v_1(x)).$$
From this we deduce that all solutions of the Equation (12) are given by
$$\vartheta(x)=v_1(x)-\frac{1}{u(x^{-1})^2}\overline{\theta}_i(v_1(x))=v_1(x)-\frac{1}{u(x^{-1})^2}v_1(x^{-1})$$
where $v_1(x)\in K_{i,1}=\overline{U}_{i,1}$, i.e., $\vartheta(x)\in \mathcal{V}_i^{(w(x))}$. As $|K_{i,1}|=p^{md_i}$, the number of left ideals $C$ of $R_i$ is equal to $p^{md_i}(p^{md_i}-1)$ in this case.

\par
   Then by Lemma 3.12(i) and Equation (11), we have that
$|C|=|\Gamma_i(\overline{w}(x)+y)|^2=p^{4md_i}$ and ${\rm w}_{H}^{(A_i)}(C)={\rm w}_{H}^{(K_i)}(\overline{(C:p)})
={\rm w}_{H}^{(K_i)}(\Gamma_i(\overline{w}(x)+y))=2$.
\hfill $\Box$

\vskip 3mm \par
   Now it is the time to prove Theorem 2.5. First, for each left ideal $C_i$ of
the ring $R_i=A_i[y;\theta_i]/\langle y^2-1\rangle$ listed by Theorems 3.8, 3.9 and 3.13 we give a generator matrix for $C_i$ as a linear code
over $A_i$ of length $2$, $0\leq i\leq r+t$.

\par
   For example, let $1\leq i\leq r$, $d_i\geq 2$ and $C_i=R_i(w(x)+y)+R_ip$. Then as an $A_i$-submodule of
$R_i$, a generator set of $C_i$ is $\{w(x)+y, y(w(x)+y), p, py\}$ where $y(w(x)+y)=1+\theta_i(w(x))y$.
By the identification of $R[D_{2n}]$ with
$\mathcal{A}^2$ under $\Xi$ defined before Theorem 2.1, we have
$$C_i=\{\xi M_i\mid \xi\in A_i^4\} \
{\rm where} \ M_i=\left[\begin{array}{cc}w(x) & 1 \cr 1 & \theta_i(w(x))\cr p & 0 \cr 0 & p\end{array}\right].$$
Since $w(x)$ is an invertible element of $A_i$ and $w(x)\cdot \theta_i(w(x))=1$, there is an invertible matrix $P$ over $A_i$
such that $PM_i=\left[\begin{array}{cc}w(x) & 1 \cr 0 & p \cr 0 & 0 \cr 0 & 0\end{array}\right]$. So $G_i=\left[\begin{array}{cc}w(x) & 1 \cr 0 & p \end{array}\right]$ is a generator matrix of $C_i$.

\par
  Now, let $r+1\leq i\leq r+t$ and $C_i=R_i\epsilon_{i,1}(x)+R_ip$. Then as an $A_i$-submodule of
$R_i$, a generator set of $C_i$ is $\{\epsilon_{i,1}(x), y\epsilon_{i,1}(x), p, py\}$ where $y\epsilon_{i,1}(x)=\theta_i(\epsilon_{i,1}(x))y=
\epsilon_{i,2}(x)y$ by Lemma 3.11(iv).
By the identification of $R[D_{2n}]$ with
$\mathcal{A}^2$ under $\Xi$ defined before Theorem 2.1, we have
$$C_i=\{\xi M_i\mid \xi\in A_i^4\} \
{\rm where} \ M_i=\left[\begin{array}{c}M_{i,1} \cr pI_2\end{array}\right].$$
As $\epsilon_{i,1}(x)\epsilon_{i,2}(x)=0$ and $\epsilon_{i,1}(x)+\epsilon_{i,2}(x)=1$ by Lemma 3.11, there is an invertible matrix $P$ over $A_i$
such that $PM_i=\left[\begin{array}{c}M_{i,1} \cr pM_{i,2}\end{array}\right]$. So $G_i=\left[\begin{array}{c}M_{i,1} \cr pM_{i,2}\end{array}\right]$ is a generator matrix of $C_i$.

\par
  Let $r+1\leq i\leq r+t$ and $C_i=R_i(\epsilon_{i,1}(x)+pb_{i,1}(x)y)$ where $b_{i,1}(x)\in K_{i,1}$.
Then as an $A_i$-submodule of
$R_i$, a generator set of $C_i$ is $\{\epsilon_{i,1}(x)+pb_{i,1}(x)y, y(\epsilon_{i,1}(x)+pb_{i,1}(x)y)\}$ where
$y(\epsilon_{i,1}(x)+pb_{i,1}(x)y)=pb_{i,1}(x^{-1})+\epsilon_{i,2}(x)y$ with $b_{i,1}(x^{-1})=\overline{\theta}_i(b_{i,1}(x))\in K_{i,2}$
by Lemma 3.11(iv). Hence a generator matrix of $C_i$ is given by
$G_i=\left[\begin{array}{cc}\epsilon_{i,1}(x) & pb_{i,1}(x) \cr pb_{i,1}(x^{-1}) & \epsilon_{i,2}(x)\end{array}\right]$.

\par
  It is routine to prove that $G_i$ is a generator matrix of $C_i$ for other cases. Here, we omit the proofs.

\vskip 3mm\par
   Finally, we prove that $\mathcal{C}^{\bot_E}=\oplus_{i=0}^{r+t}\mathcal{A}_i\Box_{\varphi_i}V_i$
as a linear $R$-code of length $2n$.
 Let $\textbf{a}=(a_{0,0}, a_{1,0},\ldots, a_{n-1,0},a_{0,1}, a_{1,1},\ldots, a_{n-1,1}),\textbf{b}=(b_{0,0}, b_{1,0},b_{2,0},
   \ldots$, $b_{n-1,0},b_{0,1}, b_{1,1},\ldots, b_{n-1,1})\in R^{2n}$. Recall that the \textit{Euclidian inner product} of $\textbf{a}$ and $\textbf{b}$ is defined by
$[\textbf{a},\textbf{b}]_E=\sum_{i=0}^{n-1}\sum_{j=0}^1a_{i,j}b_{i,j}\in R.$

\par
   Let $C$ be a linear $R$-code of length $2n$, i.e., an $R$-submodule of $R^{2n}$, the \textit{Euclidian dual code} of $C$ is defined as
$C^{\bot_E}=\{\textbf{b}\in R^{2n}\mid [\textbf{a},\textbf{b}]_E=0, \ \forall
\textbf{a}\in C\}$. Moreover, $C$ is said to be \textit{self-dual} (resp. \textit{self-orthogonal}) if
$C=C^{\bot_E}$ (resp. $C\subseteq C^{\bot_E}$).

\vskip 3mm \noindent
   {\bf Lemma 3.14} \textit{Using the notations above, denote}
$$a_j(x)=\sum_{k=0}^{n-1}a_{k,j}x^k,\ b_j(x)=\sum_{k=0}^{n-1}b_{k,j}x^k\in \mathcal{A}, \ j=0,1,$$
\textit{$\alpha=(a_0(x),a_1(x))$, $\beta=(b_0(x),b_1(x))$, and define
$\theta(\beta)=(\theta(b_0(x)),\theta(b_1(x)))$. Then $[\textbf{a},\textbf{b}]_E=0$ if}
$$\alpha\cdot (\theta(\beta))^{\rm tr}=a_0(x)\cdot \theta(b_0(x))+a_1(x)\cdot \theta(b_1(x))=0 \ {\rm in} \ \mathcal{A}.$$

\vskip 3mm \noindent
   {\bf Proof.} By $x^n=1$ in $\mathcal{A}$, it follows that
$$a_0(x)\cdot \theta(b_0(x))+a_1(x)\cdot \theta(b_1(x))=[\textit{\textbf{a}},\textit{\textbf{b}}]_E+h_1x+\ldots+h_{n-1}x^{n-1}$$
for some $h_1,\ldots,h_{n-1}\in R$. Hence $[\textit{\textbf{a}},\textit{\textbf{b}}]_E=0$ if $\alpha\cdot (\theta(\beta))^{\rm tr}=0$ in $\mathcal{A}$.
\hfill $\Box$

\vskip 3mm\par
  Now, denote $\mathcal{Q}=\oplus_{i=0}^{r+t}\mathcal{A}_i\Box_{\varphi_i}V_i$ where $V_i$ is given by Theorem 2.5 for any $i=0,1,\ldots,r+t$.
By the first part of Theorem 2.5 we have proved above, we see that $\mathcal{Q}$ is also a left $D_{2n}$-code over $R$,
$|C_i||V_i|=p^{4md_i}$ for all $i=0,1,\ldots,r$ and $|C_i||V_i|=p^{8md_i}$ for all $i=r+1,\ldots,r+t$, which implies
\begin{equation}
|\mathcal{C}||\mathcal{Q}|=\prod_{i=0}^{r+t}|C_i||V_i|=p^{4m(\sum_{i=0}^{r}d_i+2\sum_{i=r+1}^{r+t}d_i)}=|R|^{2n}
\end{equation}
by $|\mathcal{A}_i\Box_{\varphi_i}C_i|=|C_i|$ and $|\mathcal{A}_i\Box_{\varphi_i}V_i|=|V_i|$ for all $i=0,1,\ldots,r+t$.

\par
  Let $G=[g_{ij}(x)]_{k\times l}$ be a matrix over the commutative ring $A_i$ of size $k\times l$ with $g_{ij}(x)\in A_i$.
Define $\theta_i(G)=[\theta_i(g_{ij}(x))]_{k\times l}$, and denote the transpose of $G$ by $G^{{\rm tr}}$ where
$G^{{\rm tr}}=[h_{ij}(x)]_{l\times k}$ with $h_{ij}(x)=g_{ji}(x)$. First, we prove the following lemma.

\vskip 3mm\noindent
  {\bf Lemma 3.15} \textit{Let $0\leq i\leq r+t$. Then $G_i\cdot(\theta_i(H_i))^{{\rm tr}}=0$ for all pair $(G_i,H_i)$ of matrices listed in three tables of Theorem 2.5.}

\vskip 3mm\noindent
  {\bf Proof.}  Let $0\leq i\leq r$ and $d_i=1$. Then $A_i=R$ and $\theta_i={\rm id}_R$. It is routine
to check that $G_i\cdot(\theta_i(H_i))^{{\rm tr}}=G_i\cdot H_i^{{\rm tr}}=0$ for all pair $(G_i,H_i)$ of matrices listed in the first two tables
of Theorem 2.5.

\par
  Let $0\leq i\leq r$ and $d_i=2$. We consider the pair $(G_i,H_i)$ of matrices listed in the third table of Theorem 2.5.
For example, let $G_i=(w(x)(1+p\vartheta(x)),1)$ and $H_i=(-w(x)(1+p\vartheta(x)),1)$ where $w(x)\in \mathcal{W}_i$ and
$\vartheta(x)\in \mathcal{V}_i$. Then we have $w(x)\cdot \theta_i(w(x))=1$ and $\overline{\theta}_i(\vartheta(x))+\vartheta(x)=0$ by Lemma 3.6(iii) and Equation (8) in the proof of Theorem 3.9, respectively. Hence
\begin{eqnarray*}
G_i\cdot(\theta_i(H_i))^{{\rm tr}}&=&(w(x)(1+p\vartheta(x)),1)\cdot (-\theta_i(w(x))(1+p\overline{\theta}_i(\vartheta(x))),1)^{{\rm tr}}\\
 &=&-w(x)\cdot \theta_i(w(x))\cdot (1+p(\vartheta(x)+\overline{\theta}_i(\vartheta(x))))+1\\
 &=&0.
\end{eqnarray*}
It is routine to check that $G_i\cdot(\theta_i(H_i))^{{\rm tr}}=0$ for other pair $(G_i,H_i)$ of matrices
in the third table of Theorem 2.5.

\par
  Let $r+1\leq i\leq r+t$. We consider the pair $(G_i,H_i)$ of matrices listed in the fourth table of Theorem 2.5.
For example, let $G_i=(w(x)+p\vartheta(x),1)$ and $H_i=(-w(x)-p\vartheta(x),1)$ where $\vartheta(x)\in \mathcal{V}_i^{(w(x))}$ and $w(x)\in \mathcal{W}_i$. Then we have $w(x)\cdot \theta_i(w(x))=1$ and $\overline{w}(x)\overline{\theta}_i(\vartheta(x))+\overline{\theta}_i(\overline{w}(x))\vartheta(x)=0$ by the case (iv-2) and Equation (12) in the proof of Theorem 3.13, respectively. Hence
\begin{eqnarray*}
G_i\cdot(\theta_i(H_i))^{{\rm tr}}&=&(w(x)+p\vartheta(x),1)\cdot (-\theta_i(w(x))-p\overline{\theta}_i(\vartheta(x)),1)^{{\rm tr}}\\
 &=&-w(x)\cdot \theta_i(w(x))-p(\overline{w}(x)\overline{\theta}_i(\vartheta(x))+\vartheta(x)\overline{\theta}_i(\overline{w}(x)))+1\\
 &=&0.
\end{eqnarray*}
It is routine to check that $G_i\cdot(\theta_i(H_i))^{{\rm tr}}=0$ for other pair $(G_i,H_i)$ of matrices
in the fourth table of Theorem 2.5 by Lemma 3.10 (i) and (iv).
\hfill $\Box$

\vskip 3mm\par
  Now, let $\alpha\in \mathcal{C}$ and $\beta\in \mathcal{Q}$. By the identification of $R[D_{2n}]$ with
$\mathcal{A}^2$ under $\Xi$ defined before Theorem 2.1 and the first part of Theorem 2.5 we have proved,
it follows that
$$\alpha=\sum_{i=0}^{r+t}\varepsilon_i(x)\xi_iG_i\in \mathcal{A}^2 \
{\rm and} \ \beta=\sum_{i=0}^{r+t}\varepsilon_i(x)\eta_iH_i\in \mathcal{A}^2$$
for some $\xi_i\in A_i$ or $\xi_i\in A_i^2$ and $\eta_i\in A_i$ or $\eta_i\in A_i^2$.
From this, by Lemma 2.3(i), Lemma 3.3(ii) and Lemma 3.15 we deduce that
\begin{eqnarray*}
\alpha\cdot (\theta(\beta))^{{\rm tr}}&=&\left(\sum_{i=0}^{r+t}\varepsilon_i(x)\xi_iG_i\right)
 \cdot\left(\sum_{k=0}^{r+t}\varepsilon_k(x)(\theta_k(H_k))^{{\rm tr}}(\theta_k(\eta_k))^{{\rm tr}}\right)\\
 &=&\sum_{i=0}^{r+t}\varepsilon_i(x)\xi_i\left(G_i\cdot(\theta_k(H_i))^{{\rm tr}}\right)(\theta_i(\eta_i))^{{\rm tr}}\\
 &=&0.
\end{eqnarray*}
Hence $\mathcal{Q}\subseteq \mathcal{C}^{\bot_E}$ by Lemma 3.14. From this and by Equation (13), we deduce that
$\mathcal{C}^{\bot_E}=\mathcal{Q}$ as required.

%%%%%%%%%%%%%%%%%%%%%%%%%%%%%%%%%%%%%%%%%%%%%%%%%%%%%%%%%%%%%%%%%%%%%%%%%%%%%%%%%%%%%%%%%
\section{Self-dual and self-orthogonal left $D_{2n}$-codes over $R$}
\noindent
 In this section, we consider self-dual and self-orthogonal left $D_{2n}$-codes over $R=GR(p^2,m)$ where $m$ is a positive integer.
By Theorem 2.5, ${\cal C}$ is Euclidian self-dual if and only if $C_i=V_i$ for all $i=0,1,\ldots,r+t$. Then we have
\vskip 3mm \noindent
   {\bf Corollary 4.1} \textit{All distinct Euclidian self-dual
left $D_{2n}$-codes over $R$ are given by}:
${\cal C}=\sum_{i=0}^{r+t}({\cal A}_i\Box_{\varphi_i}C_i),$
\textit{where $C_i$ is a left ideal of $R_i$ satisfying one of the following two conditions}:

\par
  (i) \textit{Let $p$ be odd. If $0\leq i\leq r$, then $C_i=R_ip$. If $r+1\leq i\leq r+t$,
then $C_i=R_ip$, $R_i\epsilon_{i,j}(x)$ for $j=1,2$}.

\par
  (ii) \textit{Let $p=2$. If $0\leq i\leq r$, then $C_i=2R_i$. If $r+1\leq i\leq r+t$,
then $C_i=2R_i, \ R_i(\epsilon_{i,j}(x)+2b_{i,j}(x)y)$ where $b_{i,j}\in K_{i,j}$ and $j=1,2$}.
\par
  \textit{Therefore, the number of Euclidian self-dual
left $D_{2n}$-codes over $R$ is equal to $3^t$ when $p$ is odd; and
  the number of Euclidian self-dual
left $D_{2n}$-codes over $R$ is equal to $(2\cdot 2^{md_i}+1)^t$ when $p=2$}.

\vskip 3mm\par
   By Theorem 2.5, ${\cal C}$ is  Euclidian self-orthogonal if and only if $C_i\subseteq V_i$ for all $i=0,1,\ldots,r+t$. Hence we have

\vskip 3mm \noindent
   {\bf Corollary 4.2} \textit{All distinct self-orthogonal
left $D_{2n}$-codes over $R$ are given by}:
${\cal C}=\sum_{i=0}^{r+t}({\cal A}_i\Box_{\varphi_i}C_i),$
\textit{where $C_i$ is a left ideal of $R_i=A_i[y;\theta_i]/\langle y^2-1\rangle$
given by one of the following three cases}.

\vskip 2mm\par
   (i) \textit{Let $0\leq i\leq r$ and $d_i=1$. Then $C_i$ is one of the following two cases}:

\vskip 2mm\par
   (i-1) \textit{If $p$ is odd, $C_i=\{0\}, R_ip, R_ip(y-1), R_ip(y+1)$}.

\vskip 2mm\par
   (i-2) \textit{If $p=2$, $C_i=\{0\}, 2R_i, 2R_i(y-1)$}.

\vskip 2mm\par
   (ii) \textit{Let $0\leq i\leq r$ and $d_i\geq 2$. Then $C_i=\{0\}, R_ip, R_ip(w(x)+y)$ with $w(x)\in \mathcal{W}_i$}.

\vskip 2mm\par
   (iii) \textit{Let $r+1\leq i\leq r+t$. Then $C_i$ is one of the following two cases}:

\vskip 2mm\par
   (iii-1) \textit{Let $p$ is odd. Then $C_i=\{0\}, R_ip, R_ip\epsilon_{i,j}(x), R_ip(w(x)+y)$ with $w(x)\in \mathcal{W}_i$,
$R_i\epsilon_{i,j}(x)$ for all $j=1,2$.}

\vskip 2mm\par
   (iii-2) \textit{Let $p=2$. Then $C_i=\{0\}, 2R_i, 2R_i\epsilon_{i,j}(x), 2R_i(w(x)+y)$ with $w(x)\in \mathcal{W}_i$,
$R_i(\epsilon_{i,j}(x)+2b_{i,j}(x)y)$ where $b_{i,j}\in K_{i,j}$ and $j=1,2$.}

\vskip 2mm\par
  \textit{Therefore, the number of self-orthogonal
left $D_{2n}$-codes over $R$ is equal to
$$4^{\lambda}\prod_{d_i\geq 2, 1\leq i\leq r}(p^{\frac{md_i}{2}}+3)\prod_{r+1\leq i\leq r+t}(p^{md_i}+5)  \
{\rm if} \ p \ {\rm is} \ {\rm odd},$$
and the number of self-orthogonal
left $D_{2n}$-codes over $R$ is equal to
$$3^{\lambda}\prod_{d_i\geq 2, 1\leq i\leq r}(2^{\frac{md_i}{2}}+3)\prod_{r+1\leq i\leq r+t}(3\cdot 2^{md_i}+3) \
{\rm if} \ p=2,$$
where $\lambda=|\{i\mid d_i=1, 0\leq i\leq r\}|$}.

\vskip 3mm\par
  Finally, we consider left $D_{30}$-codes over $\mathbb{Z}_4$.
We know that
$x^{15}-1=\prod_{i=0}^3f_i(x)$, where $f_0(x)=x-1$, $f_1(x)=1+x+x^2$,
$f_2(x)=1+x+x^2+x^3+x^4$ and $f_3(x)=\rho_3(x)\rho^\ast_3(x)$ with $\rho_3(x)=1+3x+2x^2+x^4$, and $f_0(x), f_1(x), f_2(x),\rho_3(x),\rho^\ast_3(x)$
are pairwise coprime basic irreducible polynomials in $\mathbb{Z}_4[x]$. Hence
$r=2$, $t=1$, $d_0=1$, $d_1=2$ and $d_2=d_3=4$. By Corollary 2.6(ii), the number of
left $D_{30}$-codes over $\mathbb{Z}_4$, i.e., left ideals of the group ring $\mathbb{Z}_4[D_{30}]$, is equal to
\begin{eqnarray*}
\mathcal{N}_{(15,4,4)}&=&(2+5)\cdot (2^{2}+3\cdot 2^{\frac{2}{2}}+5)\cdot (2^{4}+3\cdot 2^{\frac{4}{2}}+5)\cdot(4^4+3\cdot 2^4+5)\\
  &=&7\cdot 15\cdot 33\cdot 309=1,070,685.
\end{eqnarray*}
By Corollary 4.1, the number of self-dual left $D_{30}$-codes over $\mathbb{Z}_4$ is equal to
$(2\cdot 2^4+1)^1=33$. By Corollary 4.2, the number of self-orthogonal left $D_{30}$-codes over $\mathbb{Z}_4$ is equal to
$3^1(2^{\frac{2}{2}}+3)(2^{\frac{4}{2}}+3)(3\cdot 2^4+3)=3\cdot5\cdot7\cdot 51=5355$.

\par
   For each $i=0,1,2,3$, denote
$F_i(x)=\frac{x^n-1}{f_i(x)}\in \mathbb{Z}_4[x]$. Then $F_i(x)$ and $f_i(x)$ are coprime polynomials in $\mathbb{Z}_4[x]$. We find
polynomials $u_i(x),v_i(x)\in \mathbb{Z}_4[x]$ such that
$u_i(x)F_i(x)+v_i(x)f_i(x)=1$.
In the following, let $\varepsilon_i(x)\in \mathcal{A}=\mathbb{Z}_4[x]/\langle x^{15}-1\rangle$ satisfying
$\varepsilon_i(x)\equiv u_i(x)F_i(x)=1-v_i(x)f_i(x)$ (mod $x^{15}-1$).
Precisely, we have

\par
  $\varepsilon_0(x)=3{x}^{14}+3{x}^{13}+3{x}^{12}+3{x}^{11}+3{x}^{10}+3{x}^{9}
+3{x}^{8}+3{x}^{7}+3{x}^{6}+3{x}^{5}+3{x}^{4}+3{x}^{3}+3{x}^{2}+3x+3$,

\par
  $\varepsilon_1(x)={x}^{14}+{x}^{13}+2{x}^{12}+{x}^{11}+{x}^{10}+2{x}^{9}+{x}^{8}+{x}
^{7}+2{x}^{6}+{x}^{5}+{x}^{4}+2{x}^{3}+{x}^{2}+x+2$,

\par
  $\varepsilon_2(x)={x}^{14}+{x}^{13}+{x}^{12}+{x}^{11}+{x}^{9}+{x}^{8}+{x}^{7}+{x}^{6}+{x
}^{4}+{x}^{3}+{x}^{2}+x$,

\par
  $\varepsilon_3(x)={x}^{14}+{x}^{13}+{x}^{12}+{x}^{11}+{x}^{9}+{x}^{8}+{x}^{7}+{x}^{6}+{x
}^{4}+{x}^{3}+{x}^{2}+x$.

\par
   Using the notations in Section 2 and Section 3, denote $A_i=\mathbb{Z}_4[x]/\langle f_i(x)\rangle$ for $i=0,1,2,3$.
Precisely, we have

\par
  $\diamond$ $A_0=\mathbb{Z}_4[x]/\langle x-1\rangle=\mathbb{Z}_4$ and $R_0=\mathbb{Z}_4[y]/\langle y^2-1\rangle$.

\par
  $\diamond$ $A_1=\mathbb{Z}_4[x]/\langle f_1(x)\rangle=\{a_0+a_1x\mid a_0,a_1\in \mathbb{Z}_4\}$ which is a Galois ring
of characteristic $4$ and cardinality $4^2$, and $R_1=A_1[y;\theta_1]/\langle y^2-1\rangle$ where $\theta_1(a(x))=a(x^{14})$ (mod $f_1(x)$)
for all $a(x)\in A_1$. We find an invertible element $\zeta_1(x)$ in $A_1$ of order $3$:
$\zeta_1(x)=x$. Then $\overline{\zeta}_1(x)=x$. Let

\par
  $\mathcal{W}_1=\{\zeta_1(x)^{(2-1)s}\mid s=0,1,2\}=\{1,x,3+3x\}$;

\par
  $\mathcal{V}_1=\{0\}\cup\{\overline{\zeta}_1(x)^{(2+1)\cdot 0}\}=\{0,1\}$.

\par
  $\diamond$ $A_2=\mathbb{Z}_4[x]/\langle f_2(x)\rangle=\{\sum_{j=0}^{3}a_jx^j\mid a_0,a_1,a_2,a_3\in \mathbb{Z}_4\}$  which is a Galois ring
of characteristic $4$ and cardinality $4^4$, and $R_2=A_2[y;\theta_1]/\langle y^2-1\rangle$ where $\theta_2(a(x))=a(x^{14})$ (mod $f_2(x)$)
for all $a(x)\in A_2$. We find an invertible element $\zeta_2(x)=2x^3+x+1\in A_2$ having multiplicative order $15$, which implies
$\overline{\zeta}_2(x)=1+x$. Let

\par
  $\mathcal{W}_2=\{\zeta_2(x)^{(2^{\frac{4}{2}}-1)s}\mid s=0,1,2,3,2^{\frac{4}{2}}\}=\{1, 3x^3 + 3x^2 + 3x + 3, x^3, x^2, x\}$;

\par
  $\mathcal{V}_2=\{0\}\cup\{\overline{\zeta}_2(x)^{(2^{\frac{4}{2}}+1)l}\mid l=0,1,2^{\frac{4}{2}}-2\}=\{0,1,x^3+x^2+1,x^3+x^2\}$.

\par
  $\diamond$ $A_3=\mathbb{Z}_4[x]/\langle f_3(x)\rangle=\{\sum_{j=0}^{7}a_jx^j\mid a_0,a_1,a_2,a_3,a_4,a_5,a_6,a_7\in \mathbb{Z}_4\}$ which is a principal ideal ring of cardinality $4^8$, and $R_3=A_3[y;\theta_1]/\langle y^2-1\rangle$ where $\theta_3(a(x))=a(x^{14})$ (mod $f_3(x)$)
for all $a(x)\in A_3$. We find polynomials $\phi_3(x),\psi_3(x)\in \mathbb{Z}_4[x]$ such that
$\phi_3(x)\rho_3^\ast(x)+\psi_3(x)\rho_3(x)=1$. Then
define $\epsilon_{3,1}(x), \epsilon_{3,2}(x)\in A_3$ by the following equations:
$$\epsilon_{3,1}(x)\equiv \phi_3(x)\rho_3^\ast(x), \ \epsilon_{3,2}(x)\equiv \psi_3(x)\rho_3(x) \ ({\rm mod} \ f_3(x)).$$
Precisely, we have
  $\epsilon_{3,1}(x)=3\,{x}^{7}+{x}^{5}+2\,{x}^{4}+{x}^{3}+3\,{x}^{2}+2\,x+2$ and
  $\epsilon_{3,2}(x)={x}^{7}+3\,{x}^{5}+2\,{x}^{4}+3\,{x}^{3}+{x}^{2}+2\,x+3$.
Hence $\overline{\epsilon}_{3,1}(x)=x^2+x^3+x^5+x^7$ and $\overline{\epsilon}_{3,2}(x)=1+x^2+x^3+x^5+x^7$.

\par
   Let $\Upsilon_{3,1}=\mathbb{Z}_4[x]/\langle \rho_3(x)\rangle$ which is a Galois ring
of characteristic $4$ and cardinality $4^4$. We find an element $\zeta_3(x)$ in $\Upsilon_{3,1}$ of order $15$:
$\zeta_3(x)=x^2$.
Set

\par
  $\mathcal{W}_3=\{u(x)+\frac{1}{u(x^{-1})}\mid u(x)=\epsilon_{3,1}(x)\zeta_3(x)^k, \ k=0,1,\ldots,2^4-2\}$ (mod $f_3(x)$, mod $4$);

\par
  $K_{3,1}=\{0\}\cup\{\overline{\epsilon}_{3,1}(x)\overline{\zeta}_3(x)^k\mid k=0,1,\ldots,14\}$ (mod $\overline{f}_3(x)$, mod $2$),

\par
  $K_{3,2}=\{0\}\cup\{\overline{\epsilon}_{3,2}(x)\overline{\zeta}_3(x^{-1})^k\mid k=0,1,\ldots,14\}$ (mod $\overline{f}_3(x)$, mod $2$),

\par
  $\mathcal{V}_3^{(w(x))}=\{v(x)-\frac{1}{u(x^{-1})^2}v(x^{-1})\mid v(x)\in K_{3,1}\}$  (mod $\overline{f}_3(x)$, mod $2$) for each $w(x)=u(x)+\frac{1}{u(x^{-1})}$ with $u(x)=\epsilon_{3,1}(x)\zeta_3(x)^k$ and
$0\leq k\leq 14$.

\par
  Then by Theorems 3.9, 3.10, 3.5 and 3.14, all $1070685$ distinct left $D_{30}$-codes over $\mathbb{Z}_4$ are given by:
$\mathcal{C}=\oplus_{i=0}^3(\mathcal{A}_i\Box_{\varphi_i}C_i)$, i.e.,
$$ \mathcal{C}=\sum_{i=0}^3\{(\varepsilon_i(x)c_{i,0}(x),\varepsilon_i(x)c_{i,1}(x))\mid c_{i,0}(x)+c_{i,1}(x)y\in C_i\} \
({\rm mod} \ x^{15}-1),$$
where $C_i$ is given by one of the following four cases.

\par
  $\diamondsuit$ $C_0$ is one of the following $7$ ideals of $R_0$: $\{0\}, 2R_0, 2R_0(y-1), R_0, R_0(y-1)+2R_0, R_0(y-1),
R_9((y-1)+2)$.

\par
  $\diamondsuit$ $C_1$ is one of the following $15$ ideals of $R_1$:
\begin{center}
\begin{tabular}{lllll}\hline
case &  $N_1$  &  $C_1$ (left ideals of $R_1$) & $|C_1|$ & $d_1$ \\ \hline
(1)  & $1$  & $\diamond$ $\{0\}$ & $0$  & $0$ \\
(2)  & $2$  & $\diamond$  $2^jR_1$ \ ($j=0,1$) & $2^{4(2-j)}$  & $1$ \\
(3)  & $3$  & $\diamond$  $2R_1(w(x)+y)$ \ ($w(x)\in \mathcal{G}_1$) & $2^{2}$  & $2$ \\
(4)  & $6$  & $\diamond$  $R_1(w(x)(1+2\vartheta(x))+y)$ & $2^{4}$  & $2$ \\
     &   & \ \ \  ($w(x)\in \mathcal{G}_1$, $\vartheta(x)\in \mathcal{H}_1$) &   & \\
(5)  & $3$  & $\diamond$  $R_1(w(x)+y)+2R_1$  \ ($w(x)\in \mathcal{G}_1$) & $2^{6}$  & $1$ \\
\hline
\end{tabular}
\end{center}

\par
  $\diamondsuit$ $C_2$ is one of the following $33$ ideals of $R_2$:
\begin{center}
\begin{tabular}{lllll}\hline
case &  $N_2$  &  $C_2$ (left ideals of $R_2$) & $|C_i|$ & $d_2$ \\ \hline
(1)  & $1$  & $\diamond$ $\{0\}$ & $0$  & $0$ \\
(2)  & $2$  & $\diamond$  $2^jR_2$ \ ($j=0,1$) & $2^{8(2-j)}$  & $1$ \\
(3)  & $5$  & $\diamond$  $2R_2(w(x)+y)$ \ ($w(x)\in \mathcal{G}_2$) & $2^{4}$  & $2$ \\
(4)  & $20$  & $\diamond$  $R_2(w(x)(1+2\vartheta(x))+y)$ & $2^{8}$  & $2$ \\
     &   & \ \ \  ($w(x)\in \mathcal{G}_2$, $\vartheta(x)\in \mathcal{H}_2$) &   & \\
(5)  & $5$  & $\diamond$  $R_2(w(x)+y)+2R_2$  \ ($w(x)\in \mathcal{G}_2$) & $2^{12}$  & $1$ \\
\hline
\end{tabular}
\end{center}

\par
  $\diamondsuit$ $C_3$ is one of the following $309$ ideals of $R_3$:
{\small \begin{center}
\begin{tabular}{lllll}\hline
case &  $N_3$  &  $C_3$ (left ideals of $R_3$) & $|C_3|$ & $d_3$ \\ \hline
(1)  & $1$  & $\diamond$  $\{0\}$ & $0$  & $0$ \\
(2)  & $2$  & $\diamond$  $2^jR_3$ \ ($j=0,1$) & $2^{16(2-j)}$  & $1$ \\
(3)  & $2$  & $\diamond$  $2R_3\epsilon_{3,j}(x)$ \ ($j=1,2$) & $2^{8}$  & $1$ \\
(4)  & $15$  & $\diamond$  $2R_3(w(x)+y)$ \ ($w(x)\in \mathcal{G}_3$)& $2^{8}$  & $2$ \\
(5)  & $2$  & $\diamond$  $R_3\epsilon_{3,j}(x)+2R_3$ \ ($j=1,2$) & $2^{24}$  & $1$ \\
(6)  & $32$  & $\diamond$  $R_3(\epsilon_{i,j}(x)+2b_{i,j}(x)y)$ \ ($b_{i,j}(x)\in K_{i,j}$, $j=1,2$)& $2^{16}$  & $1$ \\
(7)  & $15$  & $\diamond$  $R_3(w(x)+y)+2R_3$ \ ($w(x)\in \mathcal{G}_3$) & $2^{24}$  & $1$ \\
(8)  & $240$  & $\diamond$  $R_3(w(x)+2\vartheta(x)+y)$ & $2^{16}$  & $2$ \\
     &   & \ \ \ $(\vartheta(x)\in \mathcal{H}_3^{(w(x))}$, $w(x)\in \mathcal{G}_3$) \  &   &  \\
\hline
\end{tabular}
\end{center}}

\par
   For example, we consider the following $120$ left $D_{30}$-codes over $\mathbb{Z}_4$:
\begin{equation}
\mathcal{C}=(\mathcal{A}_1\Box_{\varphi_1}C_1)\oplus(\mathcal{A}_2\Box_{\varphi_2}C_2),
\end{equation}
where $C_1=R_1(w_1(x)(1+2\vartheta_1(x))+y)$ ($w_1(x)\in \mathcal{W}_1$, $\vartheta_1(x)\in \mathcal{V}_1$)
and  $C_2=R_2(w_2(x)(1+2\vartheta_2(x))+y)$ ($w_2(x)\in \mathcal{W}_2$, $\vartheta_2(x)\in \mathcal{V}_2$). Then
$|\mathcal{C}|=|C_1||C_2|=4^6=4096$. Specifically, a generator matrix of $\mathcal{C}$ as a linear code over $\mathbb{Z}_4$
of length $30$ is given by
\begin{eqnarray*}
G&=&\left[\begin{array}{cc}\varepsilon_1(x)w_1(x)(1+2\vartheta_1(x)) & \varepsilon_1(x) \cr x\varepsilon_1(x)w_1(x)(1+2\vartheta_1(x)) & x\varepsilon_1(x)
\cr \varepsilon_2(x)w_2(x)(1+2\vartheta_2(x)) & \varepsilon_2(x) \cr x\varepsilon_2(x)w_2(x)(1+2\vartheta_2(x)) & x\varepsilon_2(x) \cr x^2\varepsilon_2(x)w_2(x)(1+2\vartheta_2(x)) & x^2\varepsilon_2(x) \cr x^3\varepsilon_2(x)w_2(x)(1+2\vartheta_2(x)) & x^3\varepsilon_2(x)
 \end{array}\right]\ ({\rm mod} \ x^{15}-1)
=\left[\begin{array}{c}\alpha_1\cr \alpha_2 \cr \alpha_3 \cr \alpha_4 \cr \alpha_5 \cr \alpha_6\end{array}\right]
\end{eqnarray*}
where $\alpha_k\in \mathbb{Z}_4^{30}$, say. Hence
$\mathcal{C}=\{\sum_{k=1}^6a_k\alpha_6\mid a_k\in \mathbb{Z}_4, \ k=1,2,3,4,5,6\}\subseteq \mathbb{Z}_4^{30}.$
In particular, $\mathcal{C}$ is a special $2$-quasi-cyclic code over $\mathbb{Z}_4$ of length $30$ with
basic parameters $(30, 4^6; d_H(\mathcal{C}), d_L(\mathcal{C}))$, where $d_H(\mathcal{C})$ and $d_L(\mathcal{C})$ is the minimum Hamming distance and the minimum Lee distance of $\mathcal{C}$, respectively. The existing optimal parameters for $(30, 4^6, ; d_H(\mathcal{C}), d_L(\mathcal{C}))$-codes over $\mathbb{Z}_4$ in [2] are $d_H(\mathcal{C})=12$ and $d_L(\mathcal{C})=16$. Now, we have 60 left $D_{30}$-codes over $\mathbb{Z}_4$
with parameters $(30, 4^6; d_H(\mathcal{C})=12, d_L(\mathcal{C})=20)$ given by Equation (14) and the following table for $(w_1(x),\vartheta_1(x);
w_2(x),\vartheta_2(x))$:
{\small\begin{center}
\begin{tabular}{llll|llll}\hline
$w_1(x)$ &  $\vartheta_1(x)$  &  $w_2(x)$ & $\vartheta_2(x)$ & $w_1(x)$ &  $\vartheta_1(x)$  &  $w_2(x)$ & $\vartheta_2(x)$\\ \hline
$       1$&$ 0$&$                       1$&$ x^3 + x^2 + 1$ & $x$&$ 1$& $1$&$ x^3 + x^2 + 1$\\
$       1$&$ 0$&$                       1$&$     x^3 + x^2$ & $x$&$ 1$& $1$&$     x^3 + x^2$\\
$       1$&$ 0$& $w(x)$ &$ x^3 + x^2 + 1$& $x$&$ 1$& $w(x)$ &$ x^3 + x^2 + 1$\\
$       1$&$ 0$& $w(x)$ & $   x^3 + x^2$& $       x$&$ 1$& $w(x)$ &$     x^3 + x^2$\\
$       1$&$ 0$&$                     x^3$&$ x^3 + x^2 + 1$& $       x$&$ 1$&$                     x^3$&$ x^3 + x^2 + 1$\\
$       1$&$ 0$&$                     x^3$&$     x^3 + x^2$& $       x$&$ 1$&$                     x^3$&$     x^3 + x^2$\\
$       1$&$ 0$&$                     x^2$&$ x^3 + x^2 + 1$& $       x$&$ 1$&$                     x^2$&$ x^3 + x^2 + 1$\\
$       1$&$ 0$&$                     x^2$&$     x^3 + x^2$& $       x$&$ 1$&$                     x^2$&$     x^3 + x^2$\\
$       1$&$ 0$&$                       x$&$ x^3 + x^2 + 1$& $       x$&$ 1$&$                       x$&$ x^3 + x^2 + 1$\\
$       1$&$ 0$&$                       x$&$     x^3 + x^2$& $       x$&$ 1$&$                       x$&$     x^3 + x^2$\\
$       1$&$ 1$&$                       1$&$ x^3 + x^2 + 1$& $ 3x + 3$&$ 0$&$                       1$&$ x^3 + x^2 + 1$\\
$       1$&$ 1$&$                       1$&$     x^3 + x^2$& $ 3x + 3$&$ 0$&$                       1$&$     x^3 + x^2$\\
$       1$&$ 1$&$w(x)$ &$ x^3 + x^2 + 1$& $ 3x + 3$&$ 0$& $w(x)$ &$ x^3 + x^2 + 1$\\
$       1$&$ 1$&$w(x)$ & $     x^3 + x^2$& $ 3x + 3$&$ 0$& $w(x)$ &$     x^3 + x^2$\\
$       1$&$ 1$&$                     x^3$&$ x^3 + x^2 + 1$& $ 3x + 3$&$ 0$&$                     x^3$&$ x^3 + x^2 + 1$\\
$       1$&$ 1$&$                     x^3$&$     x^3 + x^2$& $ 3x + 3$&$ 0$&$                     x^3$&$     x^3 + x^2$\\
$       1$&$ 1$&$                     x^2$&$ x^3 + x^2 + 1$& $ 3x + 3$&$ 0$&$                     x^2$&$ x^3 + x^2 + 1$\\
$       1$&$ 1$&$                     x^2$&$     x^3 + x^2$& $ 3x + 3$&$ 0$&$                     x^2$&$     x^3 + x^2$\\
$       1$&$ 1$&$                       x$&$ x^3 + x^2 + 1$& $ 3x + 3$&$ 0$&$                       x$&$ x^3 + x^2 + 1$\\
$       1$&$ 1$&$                       x$&$ x^3 + x^2 $& $ 3x + 3$&$ 0$&$                       x$&$     x^3 + x^2$\\
$       x$&$ 0$&$                       1$&$ x^3 + x^2 + 1$& $ 3x + 3$&$ 1$&$                       1$&$ x^3 + x^2 + 1$\\
$       x$&$ 0$&$                       1$&$     x^3 + x^2$& $ 3x + 3$&$ 1$&$                       1$&$     x^3 + x^2$\\
$       x$&$ 0$&$w(x)$ & $ x^3 + x^2 + 1$& $ 3x + 3$&$ 1$& $w(x)$ &$ x^3 + x^2 + 1$ \\
$       x$&$ 0$&$w(x)$ & $     x^3 + x^2$& $ 3x + 3$&$ 1$& $w(x)$ &$     x^3 + x^2$\\
$       x$&$ 0$&$                     x^3$&$ x^3 + x^2 + 1$& $ 3x + 3$&$ 1$&$                     x^3$&$ x^3 + x^2 + 1$\\
$       x$&$ 0$&$                     x^3$&$     x^3 + x^2$& $ 3x + 3$&$ 1$&$                     x^3$&$     x^3 + x^2$\\
$       x$&$ 0$&$                     x^2$&$ x^3 + x^2 + 1$& $ 3x + 3$&$ 1$&$                     x^2$&$ x^3 + x^2 + 1$\\
$       x$&$ 0$&$                     x^2$&$     x^3 + x^2$& $ 3x + 3$&$ 1$&$                     x^2$&$     x^3 + x^2$\\
$       x$&$ 0$&$                       x$&$ x^3 + x^2 + 1$& $ 3x + 3$&$ 1$&$                       x$&$ x^3 + x^2 + 1$\\
$       x$&$ 0$&$                       x$&$     x^3 + x^2$& $ 3x + 3$&$ 1$&$                       x$&$     x^3 + x^2$\\
\hline
\end{tabular}
\end{center}}

\noindent
where $w(x)=3x^3 + 3x^2 + 3x + 3\in \mathcal{W}_2$.

\vskip 3mm \noindent {\bf Acknowledgments}
  Part of this work was done when Yonglin Cao was visiting Chern Institute of Mathematics, Nankai University, Tianjin, China. Yonglin Cao would like to thank the institution for the kind hospitality. This research is
supported in part by the National Natural Science Foundation of
China (Grant Nos. 11671235, 11471255, 61571243) and the National Key Basic Research Program of China (Grant No. 2013CB834204).



\end{document}